\documentclass[preprint,12pt]{elsarticle}

\usepackage{graphicx}
\usepackage{graphics}
\usepackage{amsmath}
\usepackage{amsthm}
\usepackage{amsfonts}
\usepackage{amssymb}
\usepackage[T1]{fontenc}
\usepackage{textcomp}
\usepackage{lineno}
\usepackage{lscape}
\usepackage{epstopdf}
\usepackage{subfig}
\usepackage{multirow}
\usepackage{float}
\usepackage{pifont}
\usepackage{rotating}
\usepackage{psfrag}
\DeclareGraphicsExtensions{jpg,pdf,eps}
\usepackage{wrapfig,lipsum,booktabs}
\usepackage[titletoc]{appendix}
\DeclareSymbolFont{extraup}{U}{zavm}{m}{n}
\DeclareMathSymbol{\vardiamond}{\mathalpha}{extraup}{87}
\newcommand{\acenergy}{E_{\text{ac}}}
\newcommand{\parder}[2]{\frac{\partial #1}{\partial #2}}

\newcommand{\totder}[2]{\frac{d #1}{d #2}}

\newcommand{\secder}[2]{\frac{\partial^2 #1}{\partial #2^2}}

\newcommand{\bzero}{\mathbf{0}}

\makeatletter
\newcommand{\Rmnum}[1]{\expandafter\@slowromancap\romannumeral #1@}
\makeatother

\newcommand{\bcolon}{\boldsymbol{:}}

\newcommand{\ba}{\boldsymbol{a}}

\newcommand{\bb}{\boldsymbol{b}}
\newcommand{\bc}{\boldsymbol{c}}

\newcommand{\bff}{\boldsymbol{f}}
\newcommand{\bffhat}{\hat{\bff}}

\newcommand{\bn}{\boldsymbol{n}}
\newcommand{\bp}{\boldsymbol{p}}
\newcommand{\phat}{\hat{\bp}}
\newcommand{\phatdot}{\dot{\phat}}

\newcommand{\bt}{\boldsymbol{t}}
\newcommand{\bu}{\boldsymbol{u}}
\newcommand{\bv}{\boldsymbol{v}}

\newcommand{\bx}{\boldsymbol{x}}

\newcommand{\abs}[1]{\left\lvert#1\right\rvert}

\newcommand{\bB}{\boldsymbol{B}}
\newcommand{\bC}{\boldsymbol{C}}

\newcommand{\bF}{\boldsymbol{F}}
\newcommand{\bG}{\boldsymbol{G}}

\newcommand{\bK}{\boldsymbol{K}}

\newcommand{\bM}{\boldsymbol{M}}
\newcommand{\bN}{\boldsymbol{N}}

\newcommand{\bP}{\boldsymbol{P}}

\newcommand{\bW}{\boldsymbol{W}}

\newcommand{\bgamma}{\boldsymbol{\gamma}}

\newcommand{\uhat}{\hat{\bu}}
\newcommand{\uhatdot}{\dot{\uhat}}
\newcommand{\vhat}{\hat{\bv}}
\newcommand{\bbeta}{\boldsymbol{\beta}}
\newcommand{\betahat}{\hat{\bbeta}}
\newcommand{\gammahat}{\hat{\bgamma}}

\newcommand{\bbC}{\mathbb{C}}
\newcommand{\fhat}{\hat{\bff}}

\newcommand{\del}{\boldsymbol{\nabla}}

\newcommand{\deltat}{t_{\Delta}}
\newcommand{\bcross}{\boldsymbol{\times}}

\newcommand{\btau}{\boldsymbol{\tau}}
\newcommand{\bsigma}{\boldsymbol{\sigma}}

\newcommand{\intomega}{\int_{\varOmega}}
\newcommand{\intgamma}{\int_{\varGamma}}
\newcommand{\intgammar}{\int_{\varGamma_r}}

\newcommand{\intgammainfty}{\int_{\varGamma_{\infty}}}
\newcommand{\intgammawet}{\int_{\varGamma_{\text{wet}}}}
\newcommand{\gammawet}{\varGamma_{\text{wet}}}
\newcommand{\domega}{\,d\varOmega}
\newcommand{\dgamma}{\,d\varGamma}
\newcommand{\eval}[2][\right]{\relax\ifx#1\right\relax \left.\fi#2#1\rvert}

\newcommand{\bq}{\boldsymbol{q}}
\newcommand{\br}{\boldsymbol{r}}
\newcommand{\budelta}{\bu_{\delta}}

\newcommand{\Cbar}{\bar{\bC}}

\newcommand{\bH}{\boldsymbol{H}}

\newcommand{\tbar}{\bar{\bt}}

\newcommand{\tdelta}{t_{\Delta}}

\newcommand{\beps}{\boldsymbol{\epsilon}}
\newcommand{\bepsbar}{\bar{\beps}}

\newcommand{\intgammat}{\int_{\varGamma_t}}

\newcommand{\Kwet}{\bK_{\text{wet}}}
\newcommand{\Mwet}{\bM_{\text{wet}}}
\journal{}

\begin{document}
\begin{frontmatter}



\title{Conservation properties of the trapezoidal rule in linear time domain analysis of acoustics and structures} 
\author{Arup Kumar Nandy}
\ead{arup@mecheng.iisc.ernet.in}
\author{C. S. Jog\corref{cor1}}
\ead{jogc@mecheng.iisc.ernet.in}
%\address{department of mechanical engineering, indian institute of science, bangalore, india-560012}
%\ead[url]{http://www.mecheng.iisc.ernet.in/?page_id=3917}
\address{FRITA Lab,Department of Mechanical Engineering, Indian Institute of Science, Bangalore, India-560012}
\cortext[cor1]{Corresponding author}
%%
%\author[rvt]{C.V.˜Radhakrishnan\corref{cor1}\fnref{fn1}}
%\ead{cvr@river-valley.com}
%\author[rvt,focal]{K.˜Bazargan\fnref{fn2}}
%\ead{kaveh@river-valley.com}
%\author[els]{S.˜Pepping\corref{cor2}\fnref{fn1,fn3}}
%\ead[url]{http://www.elsevier.com}
%\cortext[cor2]{Principal corresponding author}
%\fntext[fn1]{This is the specimen author footnote.}
%\fntext[fn2]{Another author footnote, but a little more longer.}
%\fntext[fn3]{Yet another author footnote. Indeed, you can have
%any number of author footnotes.}
%\address[rvt]{River Valley Technologies, SJP Building,
%Cotton Hills, Trivandrum, Kerala, India 695014}
%\address[focal]{River Valley Technologies, 9, Browns Court,
%Kennford, Exeter, United Kingdom}
%\address[els]{Central Application Management,
%Elsevier, Radarweg 29, 1043 NX\\
%Amsterdam, Netherlands}

\begin{abstract}
The trapezoidal rule, which is a special case of the Newmark family of algorithms, is one of the most widely used methods for transient hyperbolic problems. In this work, we show
that this rule conserves linear and angular momenta and energy in the case of undamped linear elastodynamics problems, and an `energy-like measure' in the case of 
undamped acoustic problems. These conservation properties, thus, provide a rational basis for using this algorithm. In linear elastodynamics problems, variants of the trapezoidal rule
that incorporate `high-frequency' dissipation are often used, since the higher frequencies, which are not approximated properly by the standard displacement-based approach, 
often result in unphysical behavior. Instead of modifying the trapezoidal algorithm, we propose using a hybrid finite element framework for constructing the stiffness matrix.
Hybrid finite elements, which are based on a two-field variational formulation involving displacement and stresses, are known to approximate the eigenvalues much more accurately
than the standard displacement-based approach, thereby either bypassing or reducing the need for high-frequency dissipation. We show this by means of several examples, where we
compare the numerical solutions obtained using the displacement-based and hybrid approaches against analytical solutions.
\end{abstract}

\end{frontmatter}

%\linenumbers

\section{Introduction}
The Newmark family of algorithms~\cite{1,2} is one of the most widely used algorithm for linear and nonlinear structural elastodynamics problems. Different strategies are
obtained by setting different values for the parameters $\beta$ and $\gamma$ in this algorithm; the trapezoidal rule is obtained by setting $\beta=0.25$ and $\gamma=0.5$. 
Simo and Tarnow~\cite{2} developed an energy-momentum conserving algorithm within the context of nonlinear elastodynamics. They showed that although the trapezoidal rule
conserves linear momentum, it fails to conserve both, the angular momentum and energy. Laursen and Chawla~\cite{TAL} and West et. al ~\cite{WKM} also represented energy conservation of Newmark algorithm in nonlinear systems. Krenk~\cite{3} shows that the trapezoidal rule conserves energy within the
context of linear elastodynamics. However, in the literature, there is no mention of whether it conserves linear and angular momenta within the context of linear
elastodynamics. In this work, we prove that besides conserving energy in linear elastodynamics and an `energy-like' measure in acoustics, 
the trapezoidal rule also conserves linear and angular momenta. Thus, although it may have shortcomings when applied to nonlinear problems, within the linear elastodynamics
context, it seems to have significant advantages.

Standard displacement-based elements (especially lower-order ones) are highly susceptible to membrane,shear and volumetric locking. Ever since the pioneering work of
Pian and Sumihara~\cite{6}, and Pian and Tong~\cite{7}, it is known that hybrid elements, which are based on a two-field variational formulation with the 
displacements and stresses interpolated separately are much less susceptible to locking than standard displacement based element, and thus, 
yield very good coarse-mesh accuracy. Hybrid elements can be used very effectively to model ``chunky'' geometries as well as beam/plate/shells with no modification in the 
mathematical formulation~\cite{8,9}. Since the stiffness matrix is approximated much more accurately, it follows that the natural frequencies are also approximated very
accurately by hybrid elements compared to standard displacement-based formulations.

There is an extensive literature on the need to incorporate `high frequency dissipation' in order to damp out the spurious participation of the higher modes~\cite{3,hugh,5,4}. 
For this purpose algorithmic damping was introduced into the Newmark algorithm. It was later realized that this also leads to low frequency damping, reducing the method to first 
order accuracy. Hilber et. al~\cite{5} introduced controllable algorithmic damping. Krenk~\cite{4} developed a state-space formulation where a fourth order accurate time 
integration algorithm with exact energy conservation is implemented. However, in this formulation, since velocity and displacements are both solved for simultaneously, it can
result in an increase in computational cost. We argue in this work that the need for high-frequency dissipation arises, not because of any inherent shortcoming of the
trapezoidal rule, but rather because of a poorly approximated stiffness matrix. In fact, as we show, similar to the continuum dynamics, the trapezoidal algorithm conserves
linear and angular momenta and energy (for undamped systems) in the absence of loading. By forming the stiffness matrix using a hybrid formulation instead of a displacement-based
one, the need for high-frequency dissipation is significantly reduced if not bypassed. We show this by comparing numerical solutions against analytical
ones obtained using either the Laplace transformation or modal methods.

The trapezoidal rule has been used extensively in the time-domain analysis of acoustical problems, for example, see Manoj and Bhattacharya~\cite{ghi} and Pinsky and Abboud~\cite{15}.
An `energy-like measure' (which does not have the units of energy) can be devised for the acoustic problem which is conserved if part of the boundary is rigid, and the 
remaining part that is radiating sound is brought to rest. For exterior acoustic problems, this energy-like measure over a finite sized (truncated) domain gradually decays as the 
sound radiation exits
the domain. We show in this work that the trapezoidal rule exactly mimics this conserving and decaying nature of the energy-like measure in interior and exterior problems, respectively.
One can thus say that the trapezoidal algorithm is an unconditionally stable algorithm for linear elastodynamics and acoustical problems in an energy sense.

For the sake of completeness, we also present a coupled formulation for structural acoustic interactions based on the trapezoidal rule that is similar to the coupled 
formulations presented by Everstine~\cite{13} and Jog~\cite{10}.

The outline of the remainder of this article is as follows. Section~\ref{secmath} presents the proofs of the conservation properties of the trapezoidal rule for linear
elastodynamics and acoustics. Section~\ref{secnumer} presents several numerical examples showing the good match between analytical and numerical solutions obtained using 
the hybrid and conventional formulations within the context of the trapezoidal rule in the case of elastodynamics and transient acoustical problems, respectively,
Finally, Section~\ref{secconcl} presents the conclusions.
\section{Conservation Properties of the Trapezoidal Rule} \label{secmath}
\subsection{Energy-momentum conserving characteristics of the trapezoidal rule in linear elastodynamics}
Let $V$ represent the domain of the structure, and let $S$ be its boundary.
Further, let $\bx$, $\bu$, $\bv$, $\beps$, $\bbC$, $\btau$, $\bt$ and $\bb$ represent the position, displacement and velocity vectors, the linearized strain tensor, the
constitutive tensor, the stress tensor, the traction vector acting on the surface, and the body force per unit mass.

The balance of linear and angular momenta, and the balance of energy within the context of linear elastodynamics are given by
\begin{gather*}
\totder{}{t}\int_V \rho\bv\,dV=\int_S \bt\,dS+\int_V\rho\bb\,dV, \\
\totder{}{t}\int_V \rho(\bx\bcross\bv)\,dV=\int_S\bx\bcross\bt\,dS+\int_V\rho\bx\bcross\bb\,dV, \\
\totder{}{t}\int_V \left[\rho\frac{\bv\cdot\bv}{2}+W(\beps)\right]\,dV=\int_S \bt\cdot\bv\,dS+\int_V\rho\bb\cdot\bv\,dV,
\end{gather*}
where $W(\beps)=\beps\bcolon\bbC\beps/2$ is the strain-energy density function.
Thus, in the absence of any displacement constraints and in the absence of loading, i.e., when $\bt=\bb=\bzero$, the linear and angular momenta, and total energy 
(i.e., kinetic+strain energy) are conserved, i.e.,
\begin{gather*}
\int_V \rho\bv\,dV=\text{constant}, \\
\int_V \rho(\bx\bcross\bv)\,dV=\text{constant}, \\
\int_V \left[\rho\frac{\bv\cdot\bv}{2}+W(\beps)\right]\,dV=\text{constant},
\end{gather*}
at all times.

The semidiscrete form (assuming $\bu=\bN_u\uhat$ and $\beps=\bB\uhat$) for the standard displacement-based formulation can be written as \cite{hugh}
\begin{equation} \label{eqham5}
\bM_s\ddot{\uhat}+\bC_s\dot{\uhat}+\bK_s\uhat=\bff_u,
\end{equation}
where $\bC_s$ is the damping matrix (that we assume to be symmetric and positive-definite), and
\begin{subequations} \label{ksmscon}
\begin{align}
\bM_s&=\int_V\rho\bN_u^T\bN_u\,dV, \\
\bK_s&=\int_V\bB_u^T\Cbar\bB_u\,dV, \\
\bff_u&=\int_V\rho\bN_u^T\bb\,dV+\int_{S_t} \bN_u^T\tbar\,dS,
\end{align}
\end{subequations}
where $S_t$ is that part of the surface over which a traction $\tbar$ is applied.

Let $\uhat_n$, $\vhat_n$ and $\fhat_{u_{n}}$ denote the nodal displacement, velocity
and load vectors, respectively, at time $t_n$, and let $\tdelta=t_{n+1}-t_n$. The trapezoidal rule for the above semi-discrete form in the absence of damping over the time-interval 
$[t_n,t_{n+1}]$ can be written as
\begin{equation} \label{eqenermom1}
\bM_s\left(\frac{\vhat_{n+1}-\vhat_n}{\tdelta}\right)+\bK_s\left(\frac{\uhat_n+\uhat_{n+1}}{2}\right)=\frac{\fhat_{u_n}+\fhat_{u_{n+1}}}{2},
\end{equation}
\begin{equation} \label{eqenermom2}
\frac{\uhat_{n+1}-\uhat_n}{\tdelta}=\frac{\vhat_n+\vhat_{n+1}}{2}.
\end{equation}
In the absence of loading, i.e., when $\fhat_{n}=\fhat_{n+1}=\bzero$, Eqn.\eqref{eqenermom1} can be written as
\begin{equation} \label{eqenermom3}
\int_V \rho \budelta\cdot\left(\frac{\bv_{n+1}-\bv_n}{\tdelta}\right)\,dV+\int_V\beps(\budelta)\bcolon\bbC\left[\frac{\beps(\bu_n)+\beps(\bu_{n+1})}{2}\right]\,dV=0\quad 
\forall\budelta.
\end{equation}

We now make special choices of $\budelta$ in Eqn.~\eqref{eqenermom3} in order to prove the conservation properties of the algorithm.
 First choose $\budelta=\bc$ over the entire
domain, where $\bc$ is a constant vector. This choice is permissible since the entire boundary is assumed to be free of displacement constraints. For this choice,
$\beps(\budelta)=\bzero$, so that we get
\begin{equation*}
\bc\cdot\int_V \rho \left(\frac{\bv_{n+1}-\bv_n}{\tdelta}\right)\,dV=0.
\end{equation*}
Since $\bc$ is arbitrary, we get
\begin{equation*}
\int_V \rho \bv_{n+1}\,dV=\int_V\rho \bv_n\,dV,
\end{equation*}
which proves that the linear momentum is conserved. 

Now choose $\budelta=\bc\bcross\bx$ over the entire domain, where $\bc$ is a constant vector. If $\bW$ is the skew tensor whose axial vector is $\bc$, then
we have $\budelta=\bW\bx$, so that $\del\budelta=\bW$, again resulting in $\beps(\budelta)=\bzero$. Using the property $(\bp\bcross\bq)\cdot\br=\bp\cdot(\bq\bcross\br)$
for vectors $\bp,\bq,\br$, Eqn.~\eqref{eqenermom3} reduces to
\begin{equation*}
\bc\cdot\int_V \rho \bx\bcross \left(\bv_{n+1}-\bv_n\right)\,dV=0,
\end{equation*}
which, by virtue of the arbitrariness of $\bc$, leads to
\begin{equation*}
\int_V \rho\bx\bcross \bv_{n+1}\,dV=\int_V\rho\bx\bcross \bv_n\,dV,
\end{equation*} 
i.e., the angular momentum is conserved.

Finally to show that the total energy is conserved, choose $\budelta=\bu_{n+1}-\bu_n$, so that $\beps(\budelta)=\beps_{n+1}-\beps_n$. Eqn.~\eqref{eqenermom3} reduces to
\begin{equation*}
\int_V\rho(\bu_{n+1}-\bu_n)\cdot\left(\frac{\bv_{n+1}-\bv_n}{\tdelta}\right)\,dV+\int_V(\beps_{n+1}-\beps_n)\bcolon\bbC\left[\frac{\beps_{n+1}+\beps_n}{2}\right]\,dV=0.
\end{equation*}
Using Eqn.~\eqref{eqenermom2}, the above equation reduces to
\begin{equation*}
\frac{1}{2}\int_V\rho(\bv_{n+1}+\bv_n)\cdot(\bv_{n+1}-\bv_n)\,dV+\frac{1}{2}\int_V(\beps_{n+1}-\beps_n)\bcolon\bbC(\beps_{n+1}+\beps_n)\,dV=0,
\end{equation*}
which, by virtue of the symmetry of the material constitutive tensor $\bbC$, reduces further to
\begin{equation} \label{eqenermom5}
\frac{1}{2}\int_V\rho\left[\bv_{n+1}\cdot\bv_{n+1}-\bv_n\cdot\bv_n\right]\,dV+
\frac{1}{2}\int_V\left[\beps_{n+1}\bcolon\bbC\beps_{n+1}-\beps_n\bcolon\bbC\beps_n\right]\,dV=0.
\end{equation}
Thus, the total energy is conserved:
\begin{equation*}
\left[\text{K.E.}+\text{Strain energy}\right]_{n+1}=\left[\text{K.E.}+\text{Strain energy}\right]_n.
\end{equation*}
Note that the choices for $\budelta$ made above belong to the finite element space for $\bu$, and hence the conservation properties hold in the fully-discrete setting.

By eliminating $\bv_{n+1}$ from Eqns.~\eqref{eqenermom1} and \eqref{eqenermom2}, we get
\begin{equation} \label{eqenermom4}
\left[\frac{2\bM_s}{\tdelta^2}+\frac{\bK_s}{2}\right]\uhat_{n+1}=\frac{2}{\tdelta^2}\bM_s\uhat_n+\frac{2}{\tdelta}\bM_s\vhat_n-\frac{1}{2}\bK_s\uhat_n+
\frac{1}{2}\left(\fhat_{u_n}+\fhat_{u_{n+1}}\right).
\end{equation}
Using the initial displacement and velocity vectors, $\uhat_0$ and $\vhat_0$, we first solve for the nodal displacement vector $\uhat_{t_1}$ at time $t_1$ using Eqn.~\eqref{eqenermom4}. 
Next, we substitute for $\uhat_{t_1}$ in Eqn.~\eqref{eqenermom2} to find $\vhat_{t_1}$. Using $\uhat_{t_1}$ and $\vhat_{t_1}$, we find $(\uhat_{t_2},\vhat_{t_2})$, and so on. 
Thus, starting from the initial displacement and velocity fields, we march forward in time, until the time instant at which the response is desired is reached.

Now consider a nonzero damping matrix $\bC_s$, which is assumed to be positive definite (note that the Rayleigh damping model $\bC_s=\alpha\bM_s+\beta\bK_s$ 
satisfies this assumption). Using
the approximation $\dot{\uhat}\approx(\uhat_{n+1}-\uhat_n)/\tdelta$ in Eqn.~\eqref{eqham5}, we now get instead of Eqn.~\eqref{eqenermom5} the relation
\begin{multline*}
\frac{1}{2}\int_V\rho\left[\bv_{n+1}\cdot\bv_{n+1}-\bv_n\cdot\bv_n\right]\,dV+
\frac{1}{2}\int_V\left[\beps_{n+1}\bcolon\bbC\beps_{n+1}-\beps_n\bcolon\bbC\beps_n\right]\,dV\\
+\frac{1}{\tdelta}(\uhat_{n+1}-\uhat_n)\cdot\bC_s(\uhat_{n+1}-\uhat_n)=0,
\end{multline*}
which by virtue of the positive definiteness of $\bC_s$ shows that
\begin{equation*}
\left[\text{K.E.}+\text{Strain energy}\right]_{n+1}\le\left[\text{K.E.}+\text{Strain energy}\right]_n.
\end{equation*}
Thus, in the presence of damping, similar to the continuum dynamics, the linear and angular momenta are conserved, and the total energy is non-increasing.
Instead of Eqn.~\eqref{eqenermom4}, we now have
\begin{multline} \label{eqenermom6}
\left[\frac{2\bM_s}{\tdelta^2}+\frac{\bC_s}{\tdelta}+\frac{\bK_s}{2}\right]\uhat_{n+1}=\frac{2}{\tdelta^2}\bM_s\uhat_n+\frac{2}{\tdelta}\bM_s\vhat_n+\frac{1}{\tdelta}\bC_s\uhat_n
-\frac{1}{2}\bK_s\uhat_n\\
+\frac{1}{2}\left(\fhat_{u_n}+\fhat_{u_{n+1}}\right).
\end{multline}
Since the energy is non-increasing, the above time-stepping scheme is unconditionally stable, i.e., there are no restrictions on the time step $\tdelta$. If $\tdelta$ is 
chosen to be a constant, then the matrix on the left hand side of Eqn.~\eqref{eqenermom6} can be decomposed right at the outset, and one merely needs to use back-substitution for
all the subsequent time steps, making the whole solution process extremely efficient.
\subsection{Hybrid element formulation for linear elastodynamics}
The hybrid formulation is based on a two-field variational principle, and is obtained by implementing both the linear momentum and the strain-displacement relations in 
weak sense \cite{8,9}, i.e., the following additional relation needs to be satisfied:
\begin{equation} \label{eqweak}
\intomega \bsigma\bcolon\left[\bepsbar(\bu)-\bbC^{-1}\btau\right]\domega=0\quad\forall\bsigma,
\end{equation}
where $\bsigma$ denotes the variation of the stress tensor $\btau$, and $\bepsbar(\bu)=[(\del\bu)+(\del\bu)^T]/2$.
Let the displacement and stress fields, and their variations, be interpolated as
\begin{align*}
\bu&=\bN_u\uhat, & \btau_c&=\bP\betahat,  \\
\bv&=\bN_u\vhat, & \bsigma_c&= \bP\gammahat,
\end{align*}
where the subscript `$c$' denotes that the engineering form (voigt notation) of that tensor. The stress shape functions $\bP$ outlined by Jog~\cite{9} are used in this work.
Defining the matrices
\begin{align*}
\bH &:= \intomega \bP^T\Cbar^{-1}\bP\domega, \\
\bG&:= \intomega \bP^T\bB_u\domega,
\end{align*}
where $\bB_u$ denotes the usual strain-displacement matrix, the stiffness matrix $\bK_s$ now takes the form
\begin{equation}\label{kshyb}
\bK_s=\bG^T\bH^{-1}\bG.
\end{equation}
The same mass matrix $\bM_s$ as in the conventional formulation is used in the hybrid formulation. Note that since the internal degrees of freedom $\betahat$
are condensed out at an element level, the number of degrees of freedom and the input data including boundary conditions are \emph{exactly the same} as in a 
conventional displacement-based formulation.  In fact, in our implementation, the same data file is used for both formulations, with just a flag indicating whether
a conventional or hybrid formulation is to be used.

The energy and momenta are conserved by the trapezoidal rule even when the stiffness matrix given by
equation~\eqref{kshyb} is used. The proof for the conservation of momenta is similar to that in the conventional formulation. In order to show energy
conservation, choose $\bsigma=(\btau_n+\btau_{n+1})/2$ in Eqn.~\eqref{eqweak} to get
\begin{align*}
\intomega \left[\frac{\btau_n+\btau_{n+1}}{2}\right]\bcolon\left[\bepsbar_n-\bbC^{-1}\btau_n\right]\domega&=0, \\
\intomega \left[\frac{\btau_n+\btau_{n+1}}{2}\right]\bcolon\left[\bepsbar_{n+1}-\bbC^{-1}\btau_{n+1}\right]\domega&=0,
\end{align*}
which leads to
\begin{align}
\intomega\left[\bepsbar_{n+1}-\bepsbar_n\right]\bcolon\left[\frac{\btau_n+\btau_{n+1}}{2}\right]\domega&=
\intomega\left[\bbC^{-1}\btau_{n+1}-\bbC^{-1}\btau_n\right]\bcolon\left[\frac{\btau_n+\btau_{n+1}}{2}\right]\domega \notag \\
&=\frac{1}{2}\intomega\left[\bbC^{-1}\btau_{n+1}\bcolon\btau_{n+1}-\bbC^{-1}\btau_n\bcolon\btau_n\right]\domega, \label{eqweak1}
\end{align}
where the last step follows from the symmetry of $\bbC$. Now we have
\begin{align*}
(\uhat_{n+1}-\uhat_n)\bK_s\left(\frac{\uhat_{n+1}+\uhat_n}{2}\right)&=(\uhat_{n+1}-\uhat_n)\bG^T\bH^{-1}\bG\left(\frac{\uhat_n+\uhat_{n+1}}{2}\right) \\
&=(\uhat_{n+1}-\uhat_n)\bG^T\left(\frac{\betahat_n+\betahat_{n+1}}{2}\right) \\
&=\intomega (\bepsbar_{n+1}-\bepsbar_n)\bcolon \left(\frac{\btau_n+\btau_{n+1}}{2}\right)\domega \\
&=\frac{1}{2}\intomega\left[\bbC^{-1}\btau_{n+1}\bcolon\btau_{n+1}-\bbC^{-1}\btau_n\bcolon\btau_n\right]\domega\quad \text{(by Eqn.~\eqref{eqweak1})} \\
&=W_{n+1}-W_n.
\end{align*}
The above term when used in place of the second term in Eqn.~\eqref{eqenermom5} proves energy conservation for the hybrid formulation.
\subsection{Energy conservation/decay in acoustical problems} \label{acou_energy}
In the previous section, we saw that for structural problems the total energy is conserved when there is no external loading and no damping. In the similar way, we now show that
an `energy-like' quantity (which has units different from that of energy) is conserved in the absence of acoustic loading. 
The wave equation is given as
\begin{equation*}
\frac{1}{a_0^2}\ddot{p}=\del^2p,
\end{equation*}
where p is acoustic pressure, and $a_0$ is the wave speed.
Multiplying the above equation by $\dot{p}$, and carrying out an appropriate integration by parts over a domain $\varOmega$ with surface $\varGamma$, we get
\begin{equation*}
\totder{}{t}\left[\intomega\left(\frac{1}{2a_0^2}\dot{p}^2+\frac{1}{2}\del p\cdot\del p\right)\domega\right]=\intgamma \dot{p}(\del p\cdot\bn)\dgamma.
\end{equation*}
Consider an interior acoustic problem, where part of the surface $\varGamma$ is rigid ($\del p\cdot\bn=0$), and where $\del p\cdot \bn$
is prescribed to have a nonzero value on the remaining part $\varGamma_r$ of the boundary. If this prescribed value is suddenly set to zero, then the `energy measure' 
$\acenergy:=\intomega\left(\frac{1}{2a_0^2}\dot{p}^2+\frac{1}{2}\del p\cdot\del p\right)\domega$ is conserved from that instant of time onwards.
Now consider an exterior acoustic problem where the domain is truncated, and let $\varGamma_{\infty}$ denote the outer surface of this truncated domain.
Since $\intgammainfty \dot{p}(\del p\cdot\bn)$ acts like a damping term, the total energy measure in the case of exterior acoustic problems is a non-increasing function
of time after the source of acoustic radiation is set to zero--if the acoustic waves have not reached $\varGamma_{\infty}$ at the instance acoustic radiation is cut off, then the
energy is conserved, while after the waves reach $\varGamma_{\infty}$, the energy decreases continuously to zero. We now show that the trapezoidal rule mimics 
this energy-conserving and energy-decaying property of the
continuum solution in the case of interior and exterior acoustic problems, respectively.
\subsection{Semi-discrete formulation of the acoustic wave equation}
The variational formulation of the wave equation is given by
\begin{equation} \label{eqvarp}
\intomega\frac{p_{\delta}}{a_0^2}\secder{p}{t}\domega+\intomega \del p_{\delta}\cdot\del p\domega=\intgamma p_{\delta}(\del p\cdot\bn)\dgamma,
\end{equation}
where $p_{\delta}$ denotes the variation of $p$. In the case of exterior problems, we take $\varGamma_{\infty}$ to be sphere of radius $R$, and 
use a spherical damper of the form
\begin{equation} \label{eqsphere}
\parder{p}{r}+\frac{1}{a_0}\parder{p}{t}=-\frac{p}{r},
\end{equation}
where $r$ denotes the spherical radial coordinate, to simulate the Sommerfeld radiation condition.  

By discretizing the pressure field and its variation as
\begin{subequations} \label{eqfemp}
\begin{align}
p&=\bN_p \phat, & p_{\delta}&=\bN_p \phat_{\delta}, \\
\del p&=\bB_p \phat, & \del p_{\delta}&=\bB_p \phat_{\delta},
\end{align}
\end{subequations}
the semi-discrete form of the finite element equations obtained using Eqn.~\eqref{eqvarp} is given by
\begin{equation}
\bM_p\ddot{\phat}+\bC_p\dot{\phat}+\bK_p\phat=\bffhat_p
\end{equation}
where, with $\ba\cdot\bn$ denoting the prescribed normal acceleration on $\varGamma_r$, and $\rho_0$ denoting the density of the acoustic fluid,
\begin{subequations} \label{kpmpcp}
\begin{align}
\bM_p&=\intomega\frac{1}{a_0^2} \bN_p^T\bN_p\domega, \\
\bC_p&=\intgammainfty \frac{1}{a_0}\bN^T_p\bN_p\dgamma, \\
\bK_p&=\intomega \bB^T_p\bB_p\domega+\intgammainfty \frac{1}{\abs{\bx}}\bN_p^T\bN_p\dgamma, \\
\bffhat_p&=-\intgammar\rho_0 (\ba\cdot\bn) \bN_p^T\dgamma.
\end{align}
\end{subequations}
\subsection{Energy conserving/decaying characteristic of the trapezoidal rule in transient acoustical problems} \label{acou_energytrap}
The trapezoidal rule for acoustical problems can be written as 
\begin{subequations} \label{eq0}
\begin{gather}
\bM_p\left(\frac{\phatdot_{n+1}-\phatdot_n}{\deltat}\right)+\bC_p\left(\frac{\phat_{n+1}-\phat_n}{\deltat}\right)+
\bK_p\left(\frac{\phat_n+\phat_{n+1}}{2}\right)=\frac{1}{2}(\fhat_{p_n}+\fhat_{p_{n+1}}), \label{eq01} \\
\frac{\phatdot_{n+1}+\phatdot_n}{2}=\frac{\phat_{n+1}-\phat_n}{\deltat}. \label{eq02}
\end{gather}
\end{subequations}
We now show that the above time-stepping scheme results in a decrease in $\acenergy$ in the absence of loads ($\bffhat_p=\bzero)$, and further, 
that it conserves $\acenergy$ if damping is also absent.

In the absence of loads, the time-stepping strategy given by Eqns.~\eqref{eq0} reduces to
\begin{subequations}
\begin{gather}
\bM_p\left(\frac{\phatdot_{n+1}-\phatdot_n}{\deltat}\right)+\bC_p\left(\frac{\phat_{n+1}-\phat_n}{\deltat}\right)+
\bK_p\left(\frac{\phat_n+\phat_{n+1}}{2}\right)=\bzero, \label{eq1} \\
\frac{\phatdot_{n+1}+\phatdot_n}{2}=\frac{\phat_{n+1}-\phat_n}{\deltat}. \label{eq2}
\end{gather}
\end{subequations}
Multiplying Eqn.~\eqref{eq1} by $\deltat(\phatdot_{n+1}+\phatdot_n)^T/2$, which is equal to $(\phat_{n+1}-\phat_n)^T$ by Eqn.~\eqref{eq2}, we get
\begin{multline} \label{eq3}
\frac{1}{2}(\phatdot_{n+1}+\phatdot_n)^T\bM_p(\phatdot_{n+1}-\phatdot_n)+(\phat_{n+1}-\phat_n)^T\bC_p\left(\frac{\phat_{n+1}-\phat_n}{\deltat}\right)\\
+\frac{1}{2}(\phat_{n+1}-\phat_n)^T\bK_p(\phat_n+\phat_{n+1})=0.
\end{multline}
Using the symmetry of $\bK_p$ and $\bM_p$, and the positive-definiteness of $\bC_p$, we get
\begin{multline*}
\left[\frac{1}{2}\phatdot_n^T\bM_p\phatdot_n+\frac{1}{2}\phat_n^T\bK_p\phat_n\right]-
\left[\frac{1}{2}\phatdot_{n+1}^T\bM_p\phatdot_{n+1}+\frac{1}{2}\phat_{n+1}^T\bK_p\phat_{n+1}\right]
=\\
(\phat_{n+1}-\phat_n)^T\bC_p\left(\frac{\phat_{n+1}-\phat_n}{\deltat}\right)\ge 0,
\end{multline*}
which shows that, similar to continuum behavior, $\acenergy$ is a non-increasing function of time. In the absence of damping (say, in interior acoustic problems with
`hard' walls), we set $\bC_p=\bzero$, so that
\begin{equation*}
\frac{1}{2}\phatdot_{n+1}^T\bM_p\phatdot_{n+1}+\frac{1}{2}\phat_{n+1}^T\bK_p\phat_{n+1}=\frac{1}{2}\phatdot_n^T\bM_p\phatdot_n+\frac{1}{2}\phat_n^T\bK_p\phat_n,
\end{equation*}
which, again, similar to the continuum behavior, implies conservation of $\acenergy$.

Eqns.~\eqref{eq0} can be written as
\begin{equation} \label{eq4}
\tilde{\bK}\phat_{n+1}=\tilde{\bff}
\end{equation}
where
\begin{align*}
\tilde{\bK}&=\frac{2}{\deltat^2}\bM_p+\frac{1}{\deltat}\bC_p+\frac{1}{2}\bK_p, \\
\tilde{\bff}&=\frac{1}{2}(\fhat_{p_n}+\fhat_{p_{n+1}})+\frac{2}{\deltat}\bM_p\phatdot_n+\frac{1}{\deltat}\bC_p\phat_n+\left(\frac{2}{\deltat^2}\bM_p
-\frac{1}{2}\bK_p\right)\phat_n, \\
\fhat_{p_{n+1}}&=-\intgammar\rho_0 (\ba\cdot\bn)_{n+1} \bN_p^T\dgamma,
\end{align*}
where $(\ba\cdot\bn)_{n+1}$ denotes the (prescribed) normal acceleration at time $t_{n+1}$.
Eqn.~\eqref{eq4} is solved for $\phat_{n+1}$, and then $\phatdot_{n+1}$ is obtained using Eqn.~\eqref{eq2}. If $\deltat$ is constant, then similar to the elastodynamics 
case, the matrix $\tilde{\bK}$ can be factored at the outset to yield a very efficient implementation. Note that there is no hybrid formulation in the acoustic case.
\subsection{Coupled formulation for transient structural acoustic interaction} 
Following Jog~\cite{10}, the semi-discrete finite element equations for the coupled structural-acoustic problem can be written as, 
\begin{align*} \label{coupled1}
\bM_s\ddot{\uhat}+\bC_s\dot{\uhat}+\bK_s\uhat+\Kwet\phat &=\bffhat_u,\\
\Mwet\ddot{\uhat}+\bM_p\ddot{\phat}+\bC_p\dot{\phat}+\bK_p\phat&=\bffhat_p,
\end{align*}
where
\begin{subequations} \label{kmwet}
\begin{align}
\Kwet&=\intgammawet\bN_u^T\bn\bN_p\dgamma, \\
\Mwet&=\intgammawet\rho_0\bN_p^T\bn^T\bN_u\dgamma.
\end{align}
\end{subequations}
$\gammawet$ is the wetted surface i.e. interface between the structure and acoustic fluid, and $\bn$ is the unit normal to this interface. Using the trapezoidal rule 
as the time stepping strategy the matrix equations for coupled structural-acoustic problems can be written as
\begin{equation}
\begin{bmatrix} \bK_{uu} & \bK_{up} \\ \bK_{pu} & \bK_{pp} \end{bmatrix} \begin{bmatrix} \uhat \\ \phat \end{bmatrix}=\begin{bmatrix} \bF_u \\ \bF_p \end{bmatrix},
\end{equation}
where
\begin{align*}
\bK_{uu}&=\frac{2}{\deltat^2}\bM_s+\frac{1}{\deltat}\bC_s+\frac{1}{2}\bK_s, \\
\bK_{up}&=\frac{1}{2}\Kwet, \\
\bK_{pu}&=\frac{2}{\deltat^2}\Mwet, \\
\bK_{pp}&=\frac{2}{\deltat^2}\bM_p+\frac{1}{\deltat}\bC_p+\frac{1}{2}\bK_p, \\
\bF_u&=\frac{1}{2}\left(\fhat_{u_n}+\fhat_{u_{n+1}}\right)+\frac{2}{\tdelta}\bM_s\vhat_n+\left(\frac{1}{\tdelta}\bC_s+\frac{2}{\tdelta^2}\bM_s
-\frac{1}{2}\bK_s\right)\uhat_n-\frac{1}{2}\Kwet\phat_n,\\
\fhat_{u_n}&=\intomega\rho\bN_u^T\bb_n\domega+\intgammat \bN_u^T\tbar_n\dgamma,\\
\bF_p&=\frac{1}{2}\left(\fhat_{p_n}+\fhat_{p_{n+1}}\right)+\frac{2}{\deltat}\bM_p\phatdot_n+\left(\frac{1}{\deltat}\bC_p+\frac{2}{\deltat^2}\bM_p-\frac{1}{2}\bK_p\right)\phat_n \\
&\quad +\frac{2}{\deltat^2}\Mwet\uhat_n+\frac{2}{\deltat}\Mwet\uhatdot_n, \\
\fhat_{p_n}&=-\intgammar\rho_0 (\ba\cdot\bn)_n \bN_p^T\dgamma.
\end{align*}
$\bK_s$ for the displacement-based and hybrid formulations is given by Eqns.~\ref{ksmscon} and \ref{kshyb}, respectively,
while $\bK_p$, $\bM_p$ and $\bC_p$ are given by Eqns.~\ref{kpmpcp}. $\Kwet$ and $\Mwet$ are given by Eqns.~\ref{kmwet}.  
Since the `acoustic energy' defined in this work does not have units of energy while the structural one does, there is no continuum conserving law for the total 
(structural+acoustic) energy in coupled structural-acoustic problems.
\section{Numerical Examples} \label{secnumer}
The good performance of the trapezoidal rule even for long-time simulations is shown by comparing the numerical solutions obtained against analytical ones.
For elastodynamics, the examples show that much superior performance is obtained using the hybrid formulation.
The analytical solutions (several of which we believe to be new) have been derived using Laplace transforms or modal methods.
SI units are used in all the examples except in cases where results are compared against works which use the FPS system. If no units are mentioned, 
then consistent SI units should be assumed.
\subsection{Acoustical Problems}
For the uncoupled acoustical problems we have assumed $\rho_0=1.2 \text{ kg/m}^3$ and $a_0=340 \text{ m/s}$.
\subsubsection{Straight duct with specified acceleration at the left end and zero acoustic pressure condition at the right side}
Consider a one-dimensional duct of length h $= 10$m. The left end is subjected to acceleration $A(t)$ and at the other end $p=0$.
For the case $A(t)=V\omega\cos\omega t$, the analytical solution is (with $k\equiv \omega/a_0$)
\begin{equation*}
p=\rho_0 a_0V\left[\frac{\sin k(h-x)}{\cos kh}\cos\omega t-\frac{2}{kh}\sum_{n=1}^{\infty}\frac{\cos\lambda_nx
\cos\lambda_na_0 t}{\left(\frac{\lambda_n}{k}\right)^2-1}\right],
\end{equation*}
where $\lambda_n=(2n-1)\pi/(2h)$. In the numerical simulations, we have taken $\omega = 200$ and $V = 2$.

\begin{figure}
 \centering
\subfloat[]{\label{tube_cos}\includegraphics[width = 16cm,height=10.0 cm]{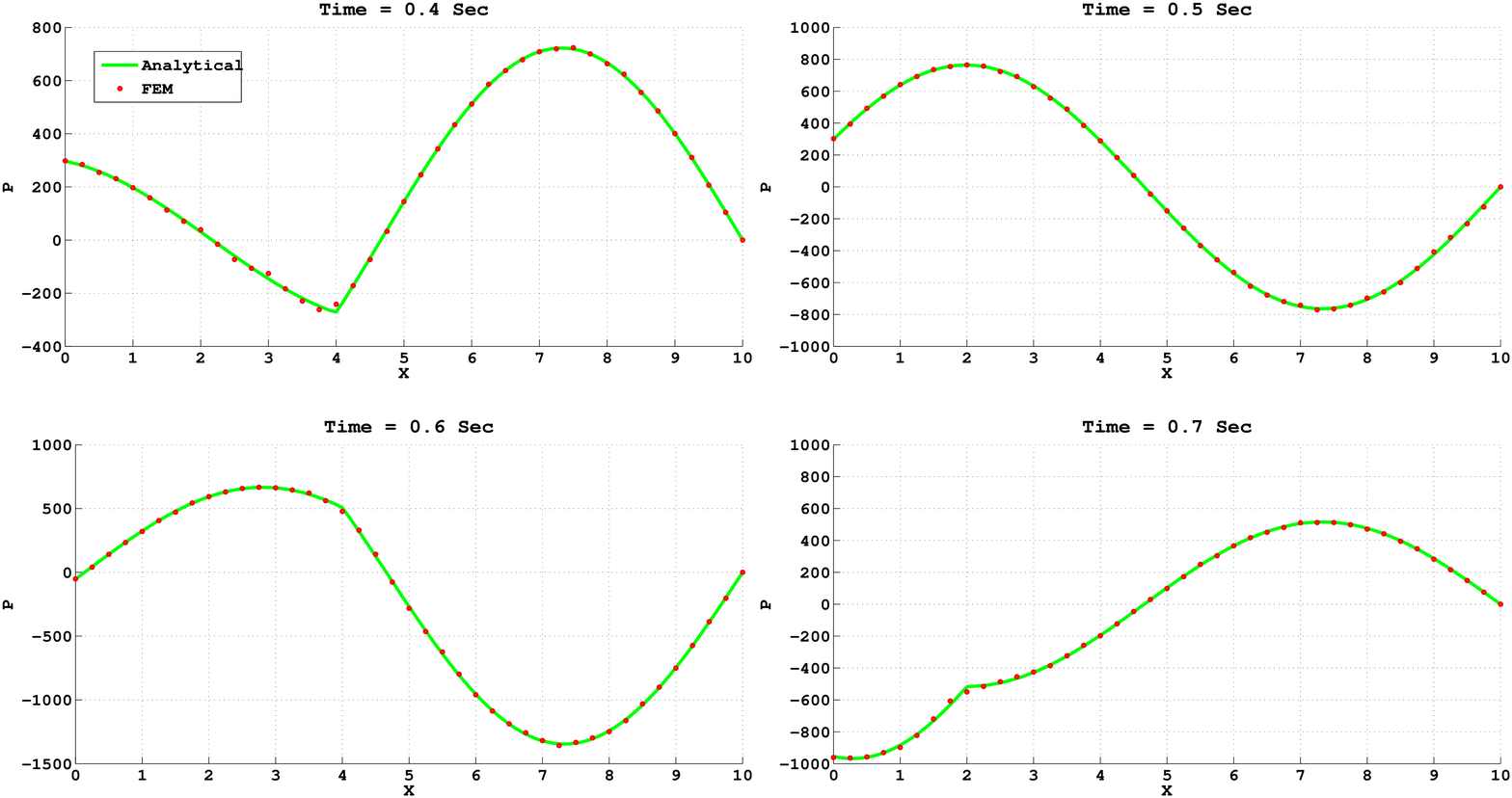}}\\
\subfloat[]{\label{tube_impulsive}\includegraphics[width = 16cm,height=10.0cm]{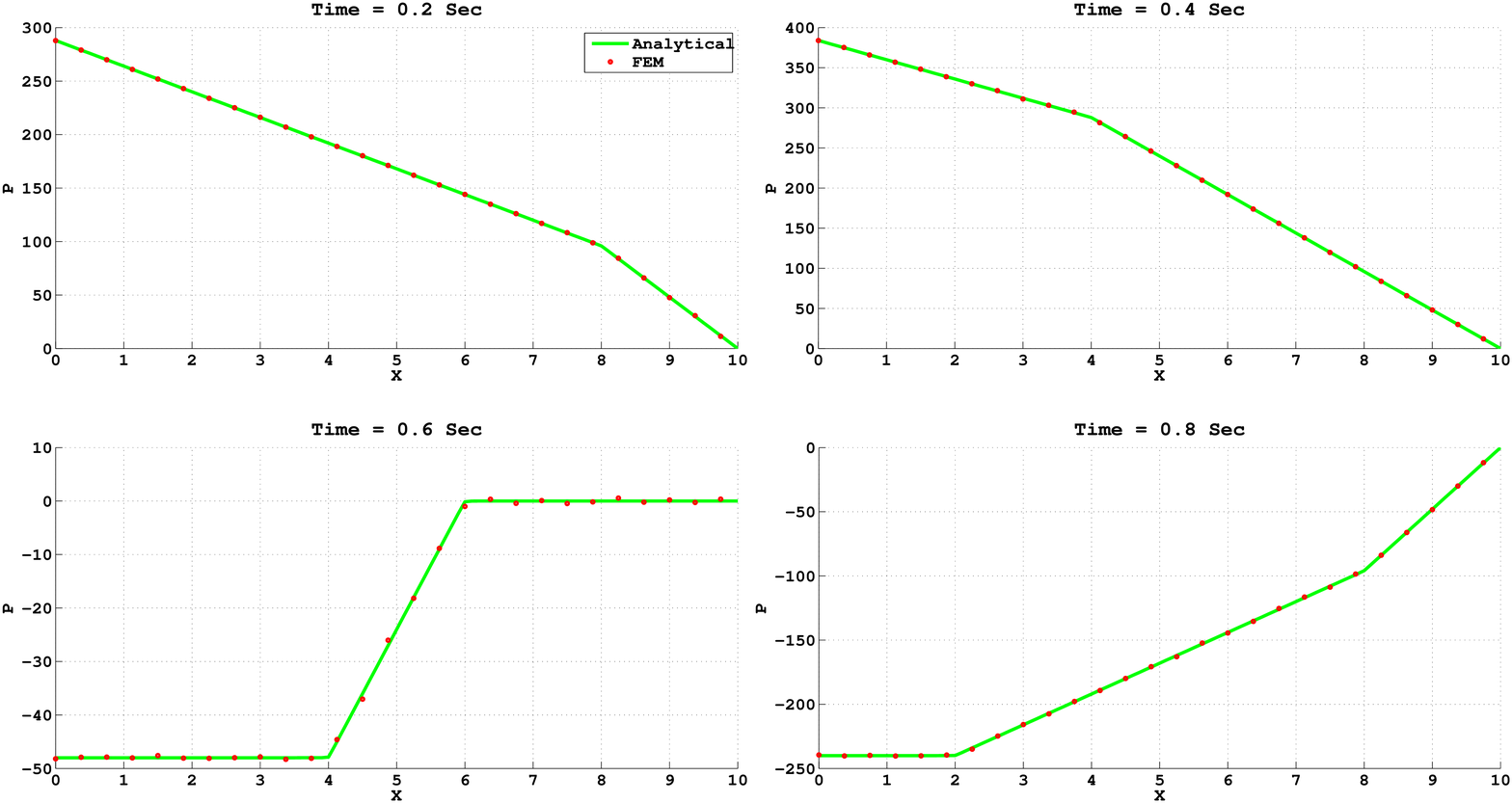}}
\caption{Pressure variation along the length of the duct at different time instants. (a) Sinusoidal Loading (b) Impulsive Loading.} \label{fig_tube}
\end{figure}

\begin{figure}
\begin{center}
\includegraphics[trim = 7cm 1cm 7cm 1cm, width = 10cm,height = 6.0cm]{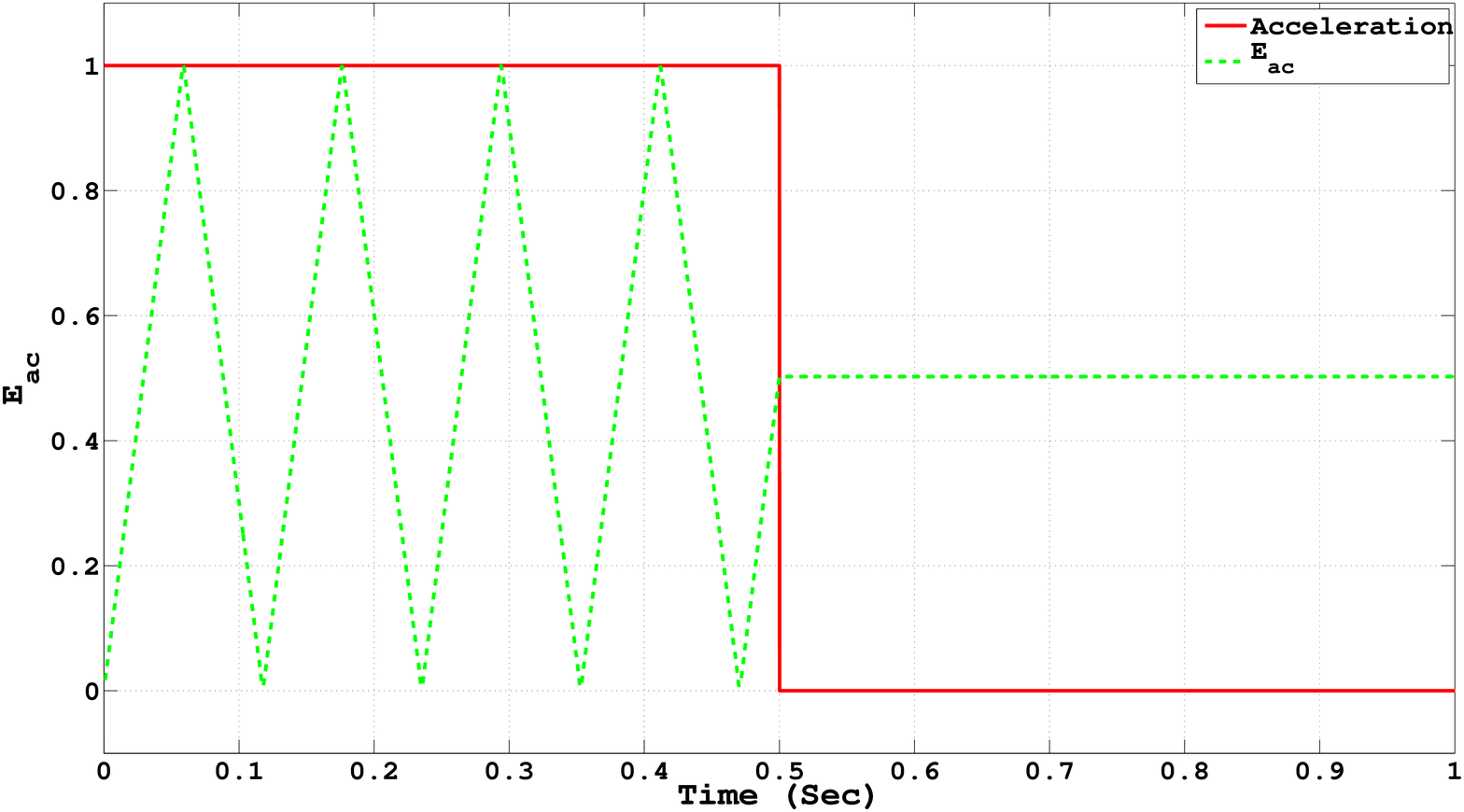}
\caption{Conservation of $\acenergy$ in the impulsive surface acceleration case in the straight duct problem.} \label{tube_impulsive_energy}
\end{center}
\end{figure}

For the case of an impulsively applied acceleration, i.e., when
\begin{equation} \label{eqimpuls}
A(t) =\begin{cases} A_0 & \text{for $t\leq t_1$}, \\ 0 & \text{for $t > t_1$}, \end{cases}
\end{equation} 
the analytical solution is
\begin{align*} \label{eqfinitetube} 
p &=\rho_0 hA_0\left[\left(1-\frac{x}{h}\right)-\sum_{n=1}^{\infty}\frac{8\cos\lambda_nx\cos\lambda_na_0t}{(2n-1)^2\pi^2}\right], && t\leq t_1 \\
&= \frac{8\rho_0 hA_0}{\pi^2}\sum_{n=1}^{\infty}\frac{\cos\lambda_nx}{(2n-1)^2}\left[\cos\lambda_na_0(t-t_1)-\cos\lambda_na_0t\right], && t>t_1.
\end{align*}
For the numerical simulations, we take $A_0 = 20$ and $t_1 = 0.5$.
The finite element results obtained using a uniform mesh of 40 quadratic elements and $\deltat = 5\times 10^{-5}$ sec are in good agreement with the analytical solution as shown in 
Figure~\ref{fig_tube}. For the impulsive surface acceleration case, since there is no driving acceleration after $t_1 = 0.5$ sec, the quantity 
$\acenergy$ should be conserved after 0.5 sec. Figure \ref{tube_impulsive_energy} (where $A(t)$ and $\acenergy$ are both normalized with
respect to their respective maximum values of $20$ and $1432$)
shows that the trapezoidal algorithm does indeed conserve the acoustic energy in accordance with the proof in Section~\ref{acou_energytrap}.
\subsubsection{Pulsating sphere in infinite domain}
Consider the case where the acoustic fluid lies outside a pulsating sphere of radius r$_1$. The normal acceleration $A(t)$ is specified at the surface r$_1$. 
The case of an explosively expanding sphere has been solved by Wu~\cite{stu} using a Kirchhoff integral formulation.
For the numerical simulations, we take $r_1=10$ m and the acoustic domain is truncated at a radius of 50 m. At this truncated surface a
spherical damper (see Eqn.~\eqref{eqsphere}) is used.

For the case when $A(t)=V_n\omega\sin\omega t$, we get $p=0$ for $t< (r-r_1)/a_0$, and for $t\ge (r-r_1)/a_0$, we get (with $k\equiv \omega/a_0$)
\begin{equation*}
p=\frac{\rho_0 a_0V_nkr_1^2}{r(1+k^2r_1^2)}\left[\sin\omega t_s-
kr_1\cos\omega t_s+kr_1e^{-\frac{a_0t_s}{r_1}}\right],
\end{equation*}
where $t_s:=t-(r-r_1)/a_0$ is the `shifted' time.
For the numerical simulations, we take $\omega = 200$ and $V_n = 2$.

When $A(t)=V_n\alpha e^{-\alpha t}$, where $\alpha$ is a positive constant, we get $p=0$ for $t< (r-r_1)/a_0$, and for $t\ge (r-r_1)/a_0$, we get
(with $k\equiv \alpha/a_0$)
\begin{equation*}
p=\frac{\rho_0 a_0 r_1^2V_nk}{r(1-kr_1)}\left[e^{-\alpha t_s}-e^{-\frac{a_0t_s}{r_1}}\right].
\end{equation*}
For the numerical simulations, we take $\alpha = 200$ and $V_n = 2$.

\begin{figure}
 \centering
\subfloat[]{\label{pulsext_sin}\includegraphics[width = 16cm,height=10.0 cm]{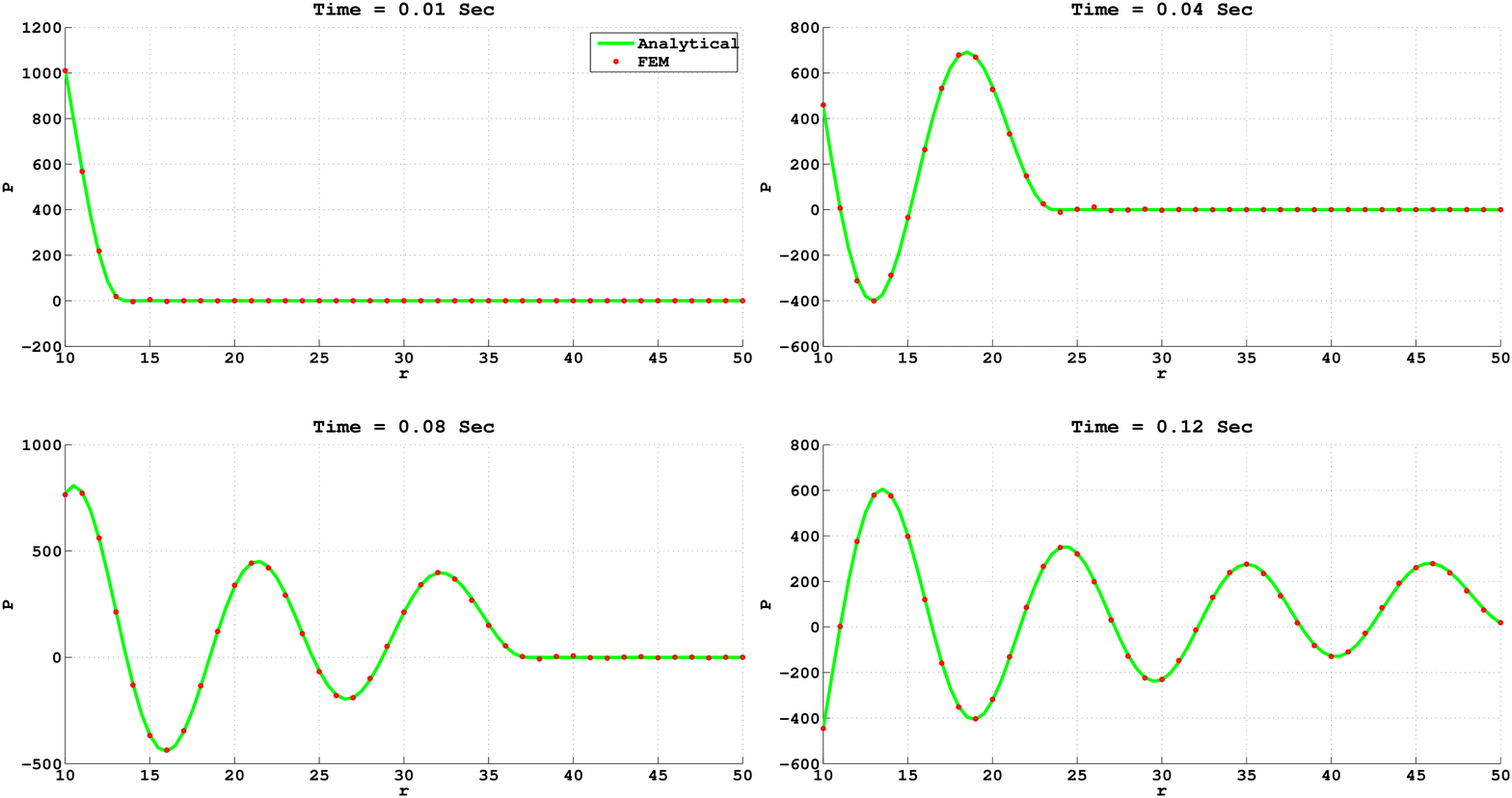}}\\
\subfloat[]{\label{pulsext_exp}\includegraphics[width = 16cm,height=10.0cm]{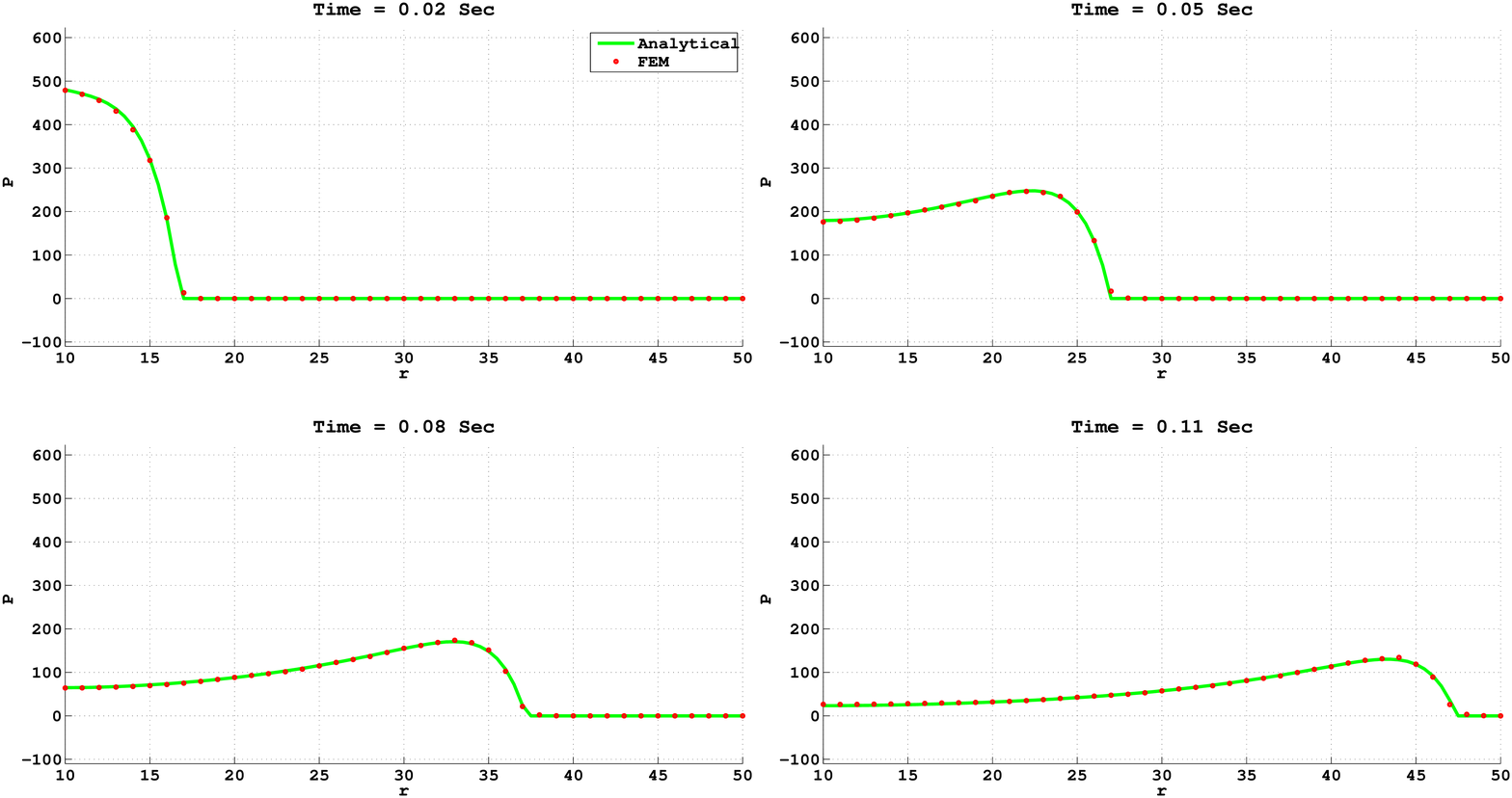}}
\caption{Pressure variation along radius at different time instants in the pulsating sphere (exterior acoustic) example. (a) Sinusoidal surface acceleration (b) Exponentially decaying surface acceleration.}
\label{fig_puls_ext}
\end{figure}

\begin{figure}
\begin{center}
\includegraphics[trim = 7cm 1cm 7cm 1cm, width = 10cm, height = 6cm]{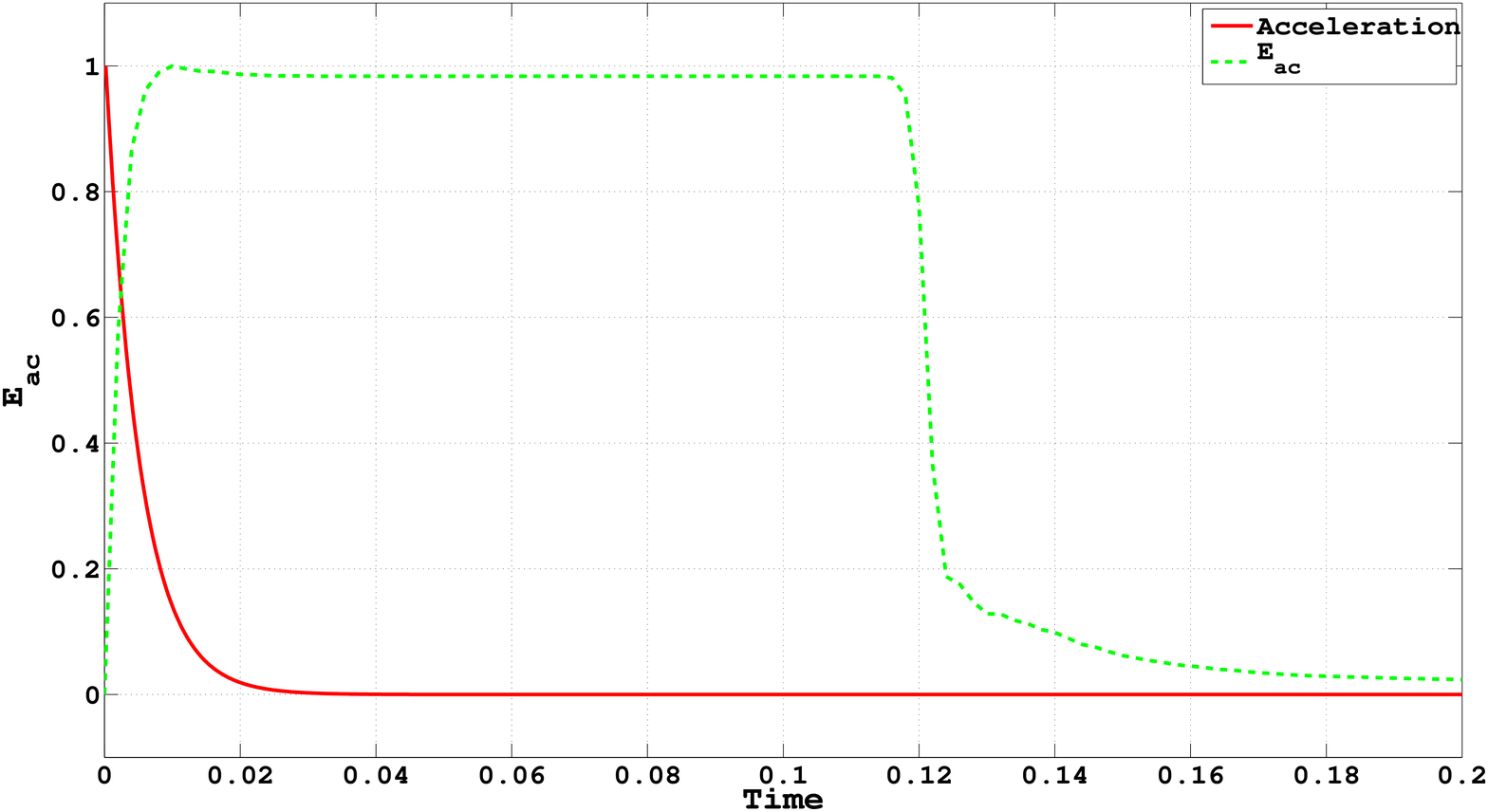}
\caption{Variation of $\acenergy$ in the exponential loading case in the pulsating sphere (exterior acoustic) problem.} \label{pulsext_exp_energy}
\end{center}
\end{figure}

As the problem is radially symmetric, we use uniform axisymmetric meshes of $n_r\times n_{\theta}=20\times 4$ and $80\times 4$ nine-node elements to model a quarter
of the cross-section and times steps of $10^{-3}$ and $5\times 10^{-4}$ s in the sinusoidal and exponentially decaying surface acceleration cases, respectively.
Figure~\ref{fig_puls_ext} again shows the good agreement between the analytical and the finite element solutions.
In the exponentially decaying surface acceleration case, the surface acceleration is almost zero after around 0.025 s (see Figure~\ref{pulsext_exp_energy}
where the acceleration and $\acenergy$ are normalized with respect to their respective maximum values of 384.32 and 17.03$\times 10^6$ over the entire time span). 
Hence from this time (0.025 s) to the time taken by the acoustic wave to reach $\varGamma_\infty$, namely $(50-10)/340=0.1176$ s, $\acenergy$ should be constant, after which
it should decay to zero as the waves exit $\varGamma_\infty$. From Figure \ref{pulsext_exp_energy}, we see that the trapezoidal algorithm simulates this conservation and 
decay of the acoustic energy very accurately.
\subsubsection{Pulsating sphere with acoustic domain inside the sphere}
Consider again a pulsating sphere of radius $r_1$, with the difference that now the acoustic fluid is inside the sphere.
Let $\lambda_n, n=1,2,\cdots,\infty,$ be the positive roots of $\tan r_1x=r_1x$.
For $A(t)=V\omega\sin\omega t$, we get (with $k\equiv \omega/a_0$)
\begin{equation*}
p=\rho_0 a_0V\left[-\frac{3\omega t}{kr_1}-\frac{kr_1^2\sin kr}{r(kr_1\cos kr_1-\sin kr_1)}\sin \omega t
+\sum_{n=1}^{\infty} \frac{2\lambda_nd_n\sin\lambda_na_0t}{\left(\frac{\lambda_n}{k}\right)^2-1}\right],
\end{equation*}
where 
\begin{equation*}
d_n=\frac{\sin\lambda_nr}{\lambda_n^2r\sin\lambda_nr_1}.
\end{equation*}
Note that $\lim_{r\rightarrow 0} d_n=1/(\lambda_n\sin\lambda_nr_1)$. For generating the numerical results, we have used $r_1=5$, $V=2$ and $\omega =200$.

\begin{figure}
 \centering
\subfloat[]{\label{puls_int_sin}\includegraphics[width = 16cm,height=10.0 cm]{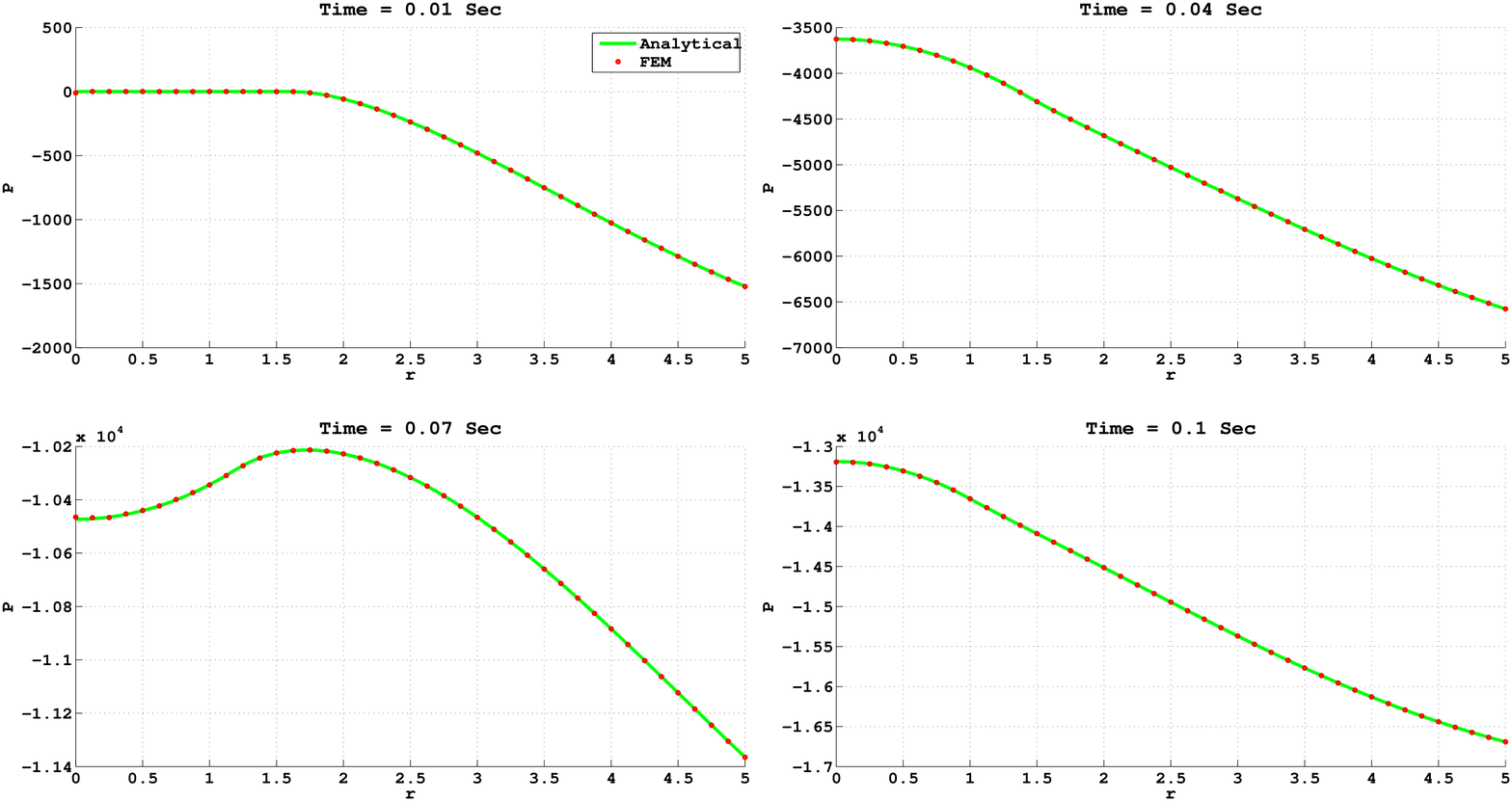}}\\
\subfloat[]{\label{puls_int_imp}\includegraphics[width = 16cm,height=10.0cm]{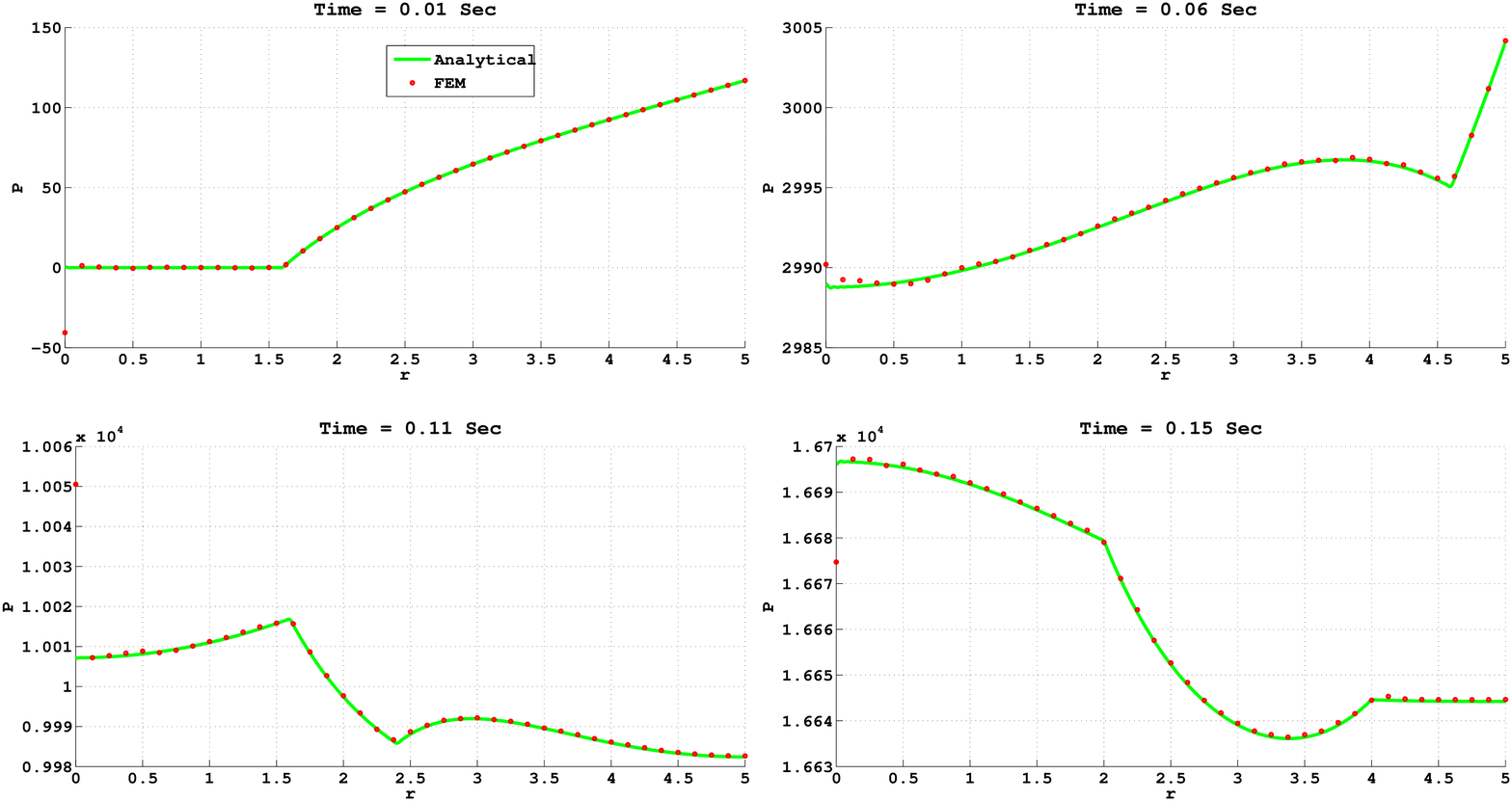}}
\caption{Pressure variation along radius for different time instants in the pulsating sphere (interior acoustic) problem. (a) Sinusoidal Loading (b) Impulsive Loading.} \label{fig_puls_int}
\end{figure}

\begin{figure}
\begin{center}
\includegraphics[trim = 7cm 1cm 7cm 1cm, width = 10cm, height = 6cm]{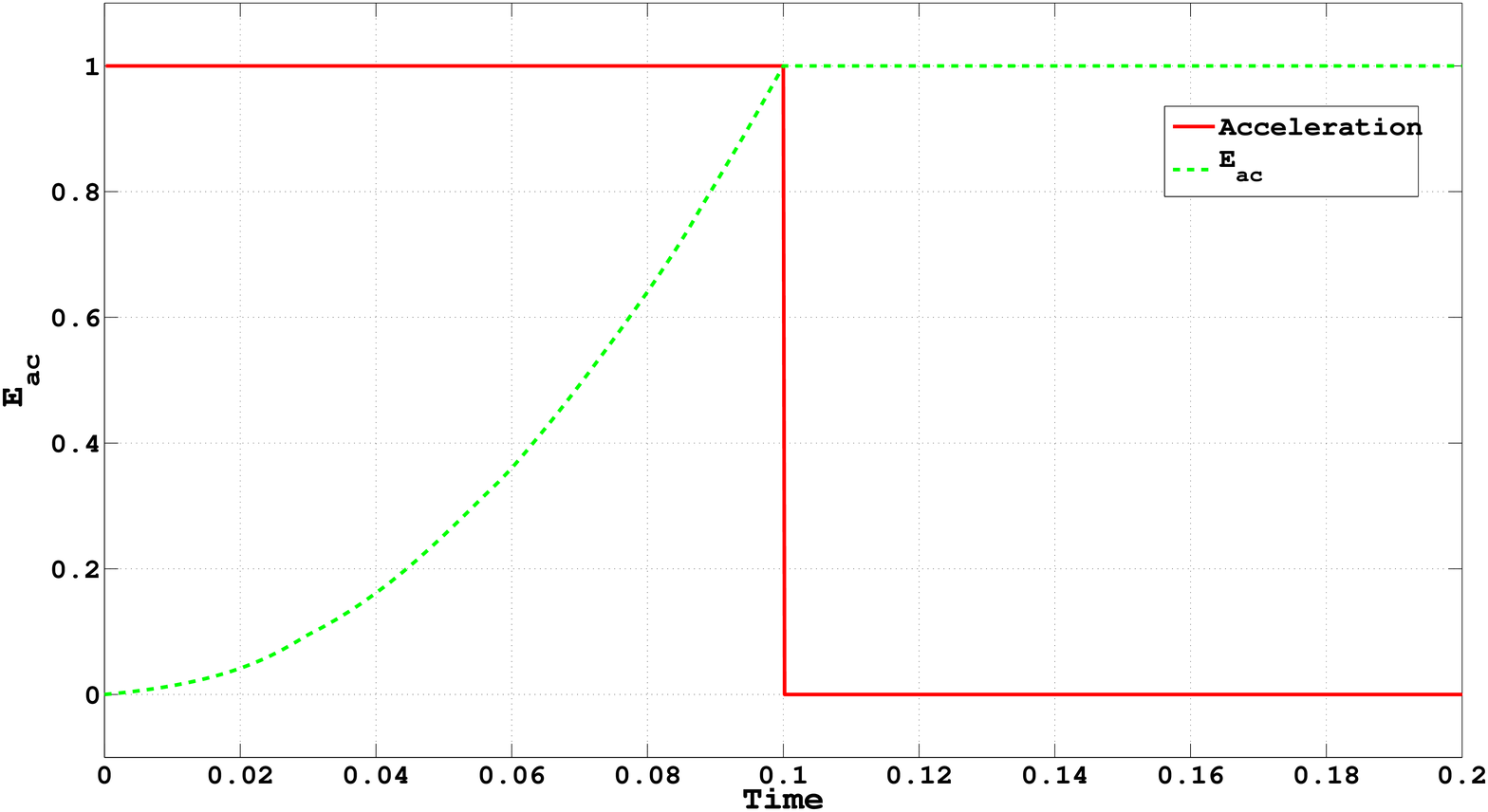}
\caption{Variation of $\acenergy$ in the pulsating sphere (interior acoustic) problem.} \label{pulsint_energy}
\end{center}
\end{figure}

For the case of an impulsively applied acceleration as given by Eqn.~\eqref{eqimpuls}, we get
\begin{align*}
p&=\rho_0 A_0\left[\frac{3r_1}{10}-\frac{r^2+3a_0^2t^2}{2r_1}+\sum_{n=1}^{\infty}2d_n\cos \lambda_na_0 t\right]
&& t\leq t_1 \\
&=-\frac{3\rho_0 a_0^2A_0t_1}{r_1}\left(t-\frac{t_1}{2}\right)-\sum_{n=1}^{\infty}2\rho_0 A_0d_n\left[\cos\lambda_na_0(t-t_1)-\cos\lambda_na_0t\right], && t>t_1.
\end{align*}
For the numerical simulations, we assume $A_0=20$ and $t_1=0.1$.
We use uniform meshes of $n_r\times n_{\theta}=40\times 4$ and $80\times 4$ nine-node axisymmetric elements to mesh a quarter of the cross-section, and
$\deltat$ of $5\times 10^{-4}$ and $2.5\times 10^{-5}$ in the sinusoidal and impulsively applied surface accelerations, respectively.
Figure \ref{fig_puls_int}
presents the analytical and the finite element pressure distributions. The conservation of $\acenergy$ for this interior acoustic problem can be seen in Figure~\ref{pulsint_energy}
where the energy has been normalized using the maximum value of $5.005\times10^6$.  
\subsubsection{Oscillating sphere in infinite domain}
\begin{figure}
\begin{center}
\includegraphics[ width = 4.5cm, height = 3.75cm]{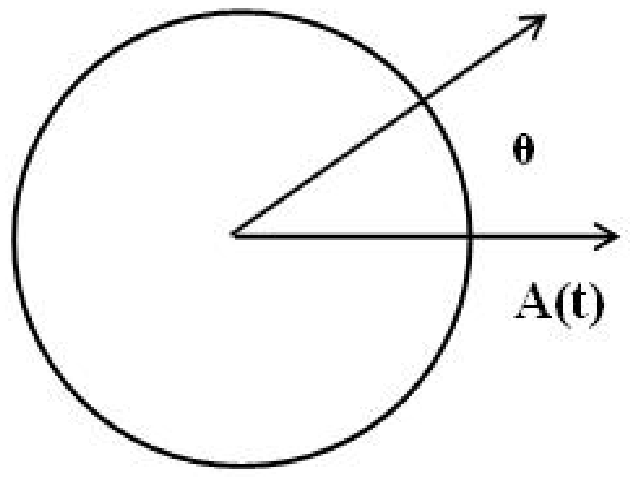}
\caption{Oscillating Sphere.} \label{oscillating}
\end{center}
\end{figure}

\begin{figure}
\centering
\subfloat[]{\label{osc_ext_135}\includegraphics[width = 16cm,height=10.0 cm]{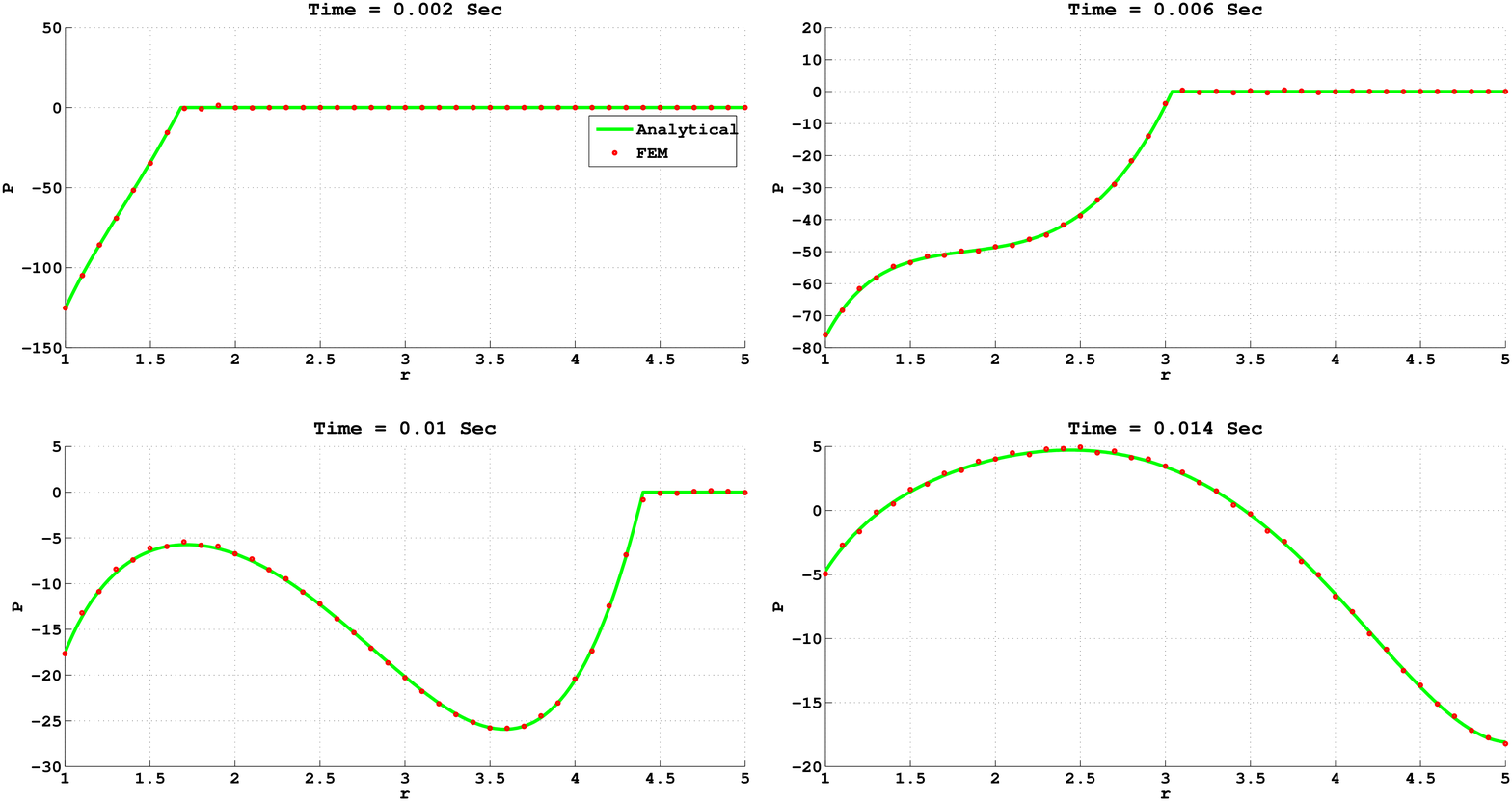}}\\
\subfloat[]{\label{osc_ext_180}\includegraphics[width = 16cm,height=10.0cm]{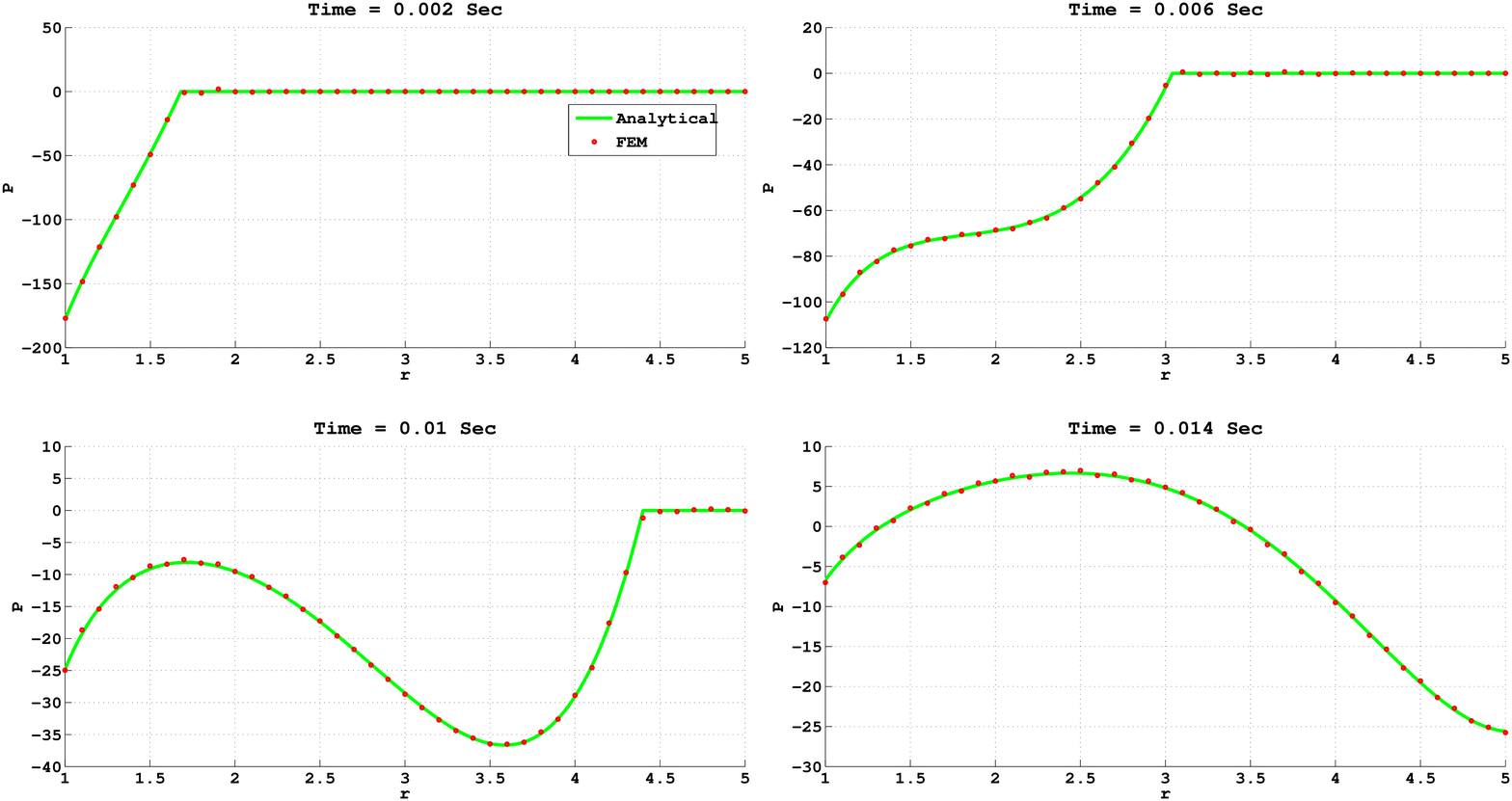}}
\caption{Pressure variation along radius for different times in the oscillating sphere (exterior acoustic) problem. (a) $\theta$ = 135 degree (b) $\theta$ = 180 degree.}
\label{fig_osc_ext}
\end{figure}

\begin{figure}
\begin{center}
\includegraphics[trim = 7cm 1cm 7cm 1cm, width = 10cm, height = 6cm]{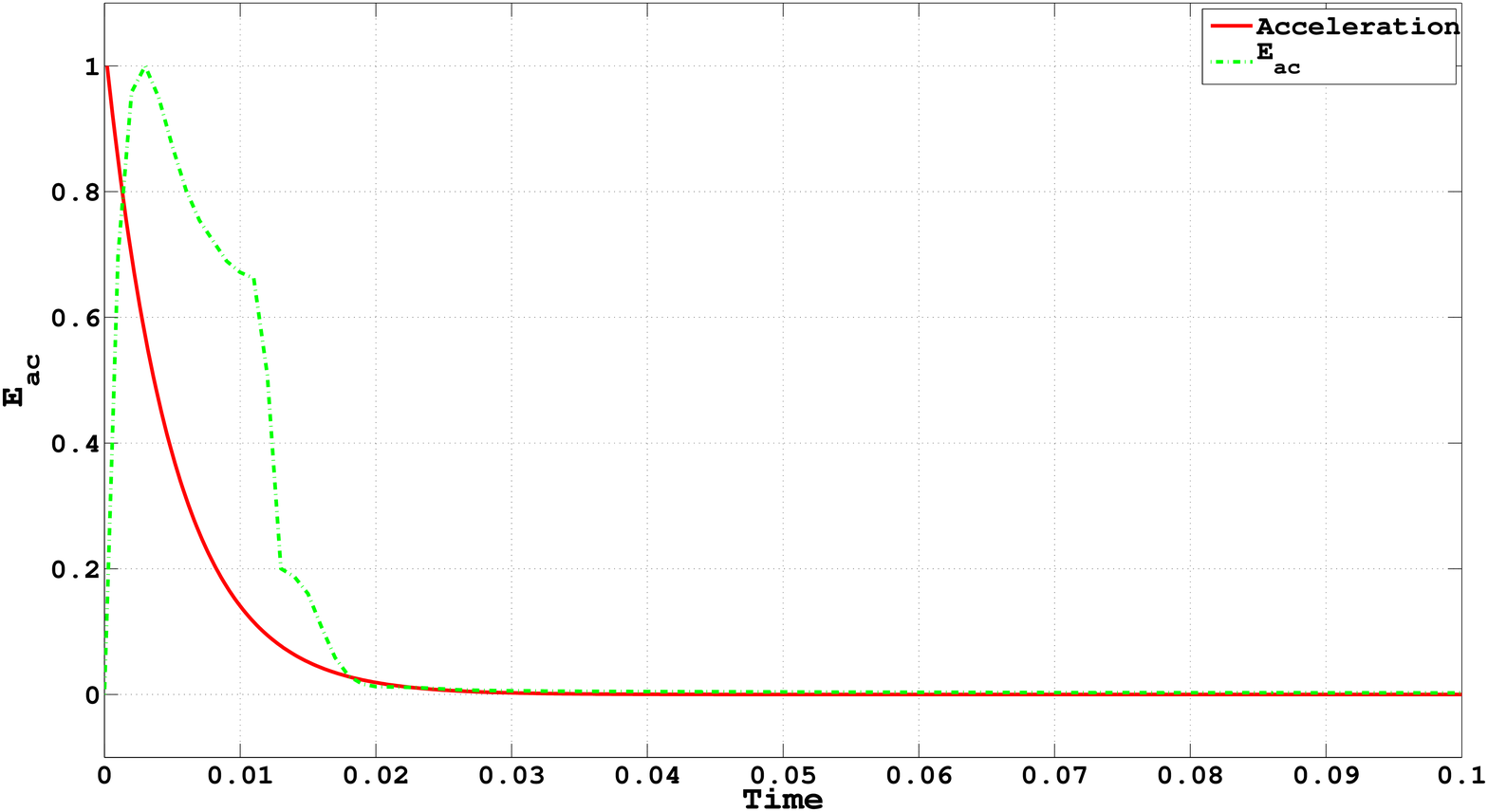}
\caption{Variation of $\acenergy$ in the oscillating sphere (exterior acoustic) problem.} \label{osc_ext_energy}
\end{center}
\end{figure}

Consider a rigid sphere of radius r$_1$ in an unbounded domain which is subjected to an acceleration A(t) in the horizontal direction as shown in Figure~\ref{oscillating}. 
The case of an impulsively accelerated rigid sphere has been solved by Wu~\cite{stu}.
When $A(t)=V\alpha e^{-\alpha t}$, where $\alpha$ is a positive constant, we get $p=0$ for $t< (r-r_1)/a_0$, and for $t\ge (r-r_1)/a_0$, we get
%\begin{align*}
%p&=\frac{\rho_0 a_0\alpha r_1^2V\cos\theta}{r^2(2a_0^2-2a_0\alpha r_1+\alpha^2r_1^2)}\Biggl[e^{-\alpha t_s}(a_0-\alpha r)r_1+
%e^{-\frac{a_0t_s}{r_1}} \\
%&\quad \left[(\alpha r-a_0)r_1\cos\frac{a_0t_s}{r_1}+[2a_0r-(a_0+\alpha r)r_1+
%\alpha r_1^2]\sin\frac{a_0t_s}{r_1}\right]\Biggr]. 
%\end{align*}
\begin{align*}
p&=\frac{\rho_0 a_0\alpha r_1^2V\cos\theta}{r^2(2a_0^2-2a_0\alpha r_1+\alpha^2r_1^2)}\Biggl[e^{-\alpha t_s}(a_0-\alpha r)r_1+
e^{-\frac{a_0t_s}{r_1}} \\
 &\left[(\alpha r-a_0)r_1\cos\frac{a_0t_s}{r_1}
 +[2a_0r-(a_0+\alpha r)r_1+\alpha r_1^2]\sin\frac{a_0t_s}{r_1}\right]\Biggr]. 
\end{align*}
For the numerical simulation, we truncate the domain at a radius of 5 m. A uniform mesh of $n_r\times n_{\theta}=40\times20$ axisymmetric nine-node elements is used to
model the semicircular region between the rigid sphere and the truncating sphere. The time step used is $\deltat = 5\times 10^{-5}$ s. V and $\alpha$ are taken as 2 and 
200, respectively. Since the solution varies along $\theta$, we compare the analytical and finite element solutions at two different $\theta$ values as shown
in Figure~ \ref{fig_osc_ext}. The acoustic energy obtained using the trapezoidal rule has the expected behavior as seen in Figure \ref{osc_ext_energy},
where the variations of $A(t)$ and $\acenergy$ are shown normalized against their maximum values of 384.32 and 51210, respectively.
\subsubsection{Oscillating sphere inside a rigid sphere}
\begin{figure}
 \centering
\subfloat[]{\label{osc_ann_135}\includegraphics[width = 16cm,height=10 cm]{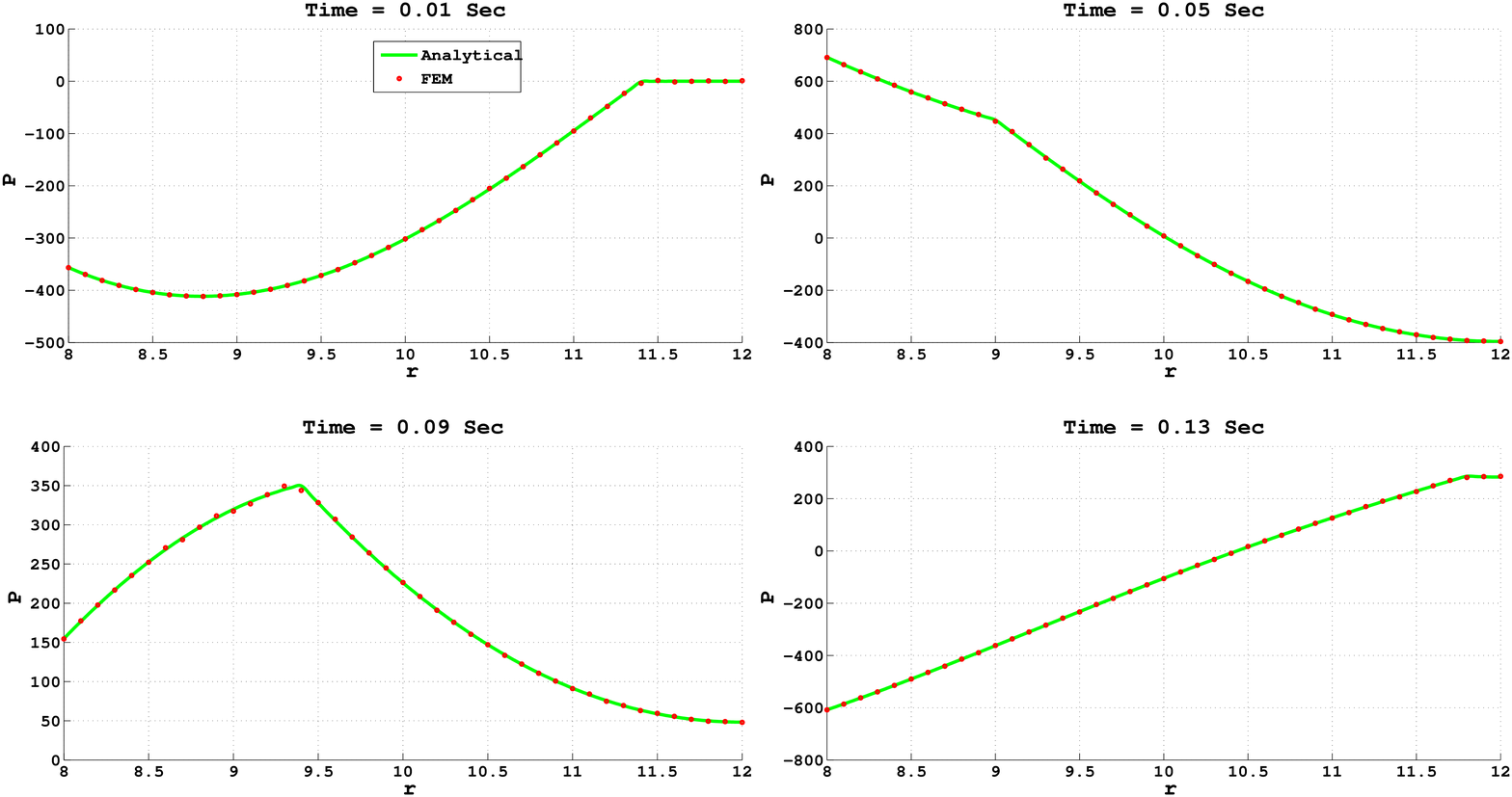}}\\
\subfloat[]{\label{osc_ann_180}\includegraphics[width = 16cm,height=10 cm]{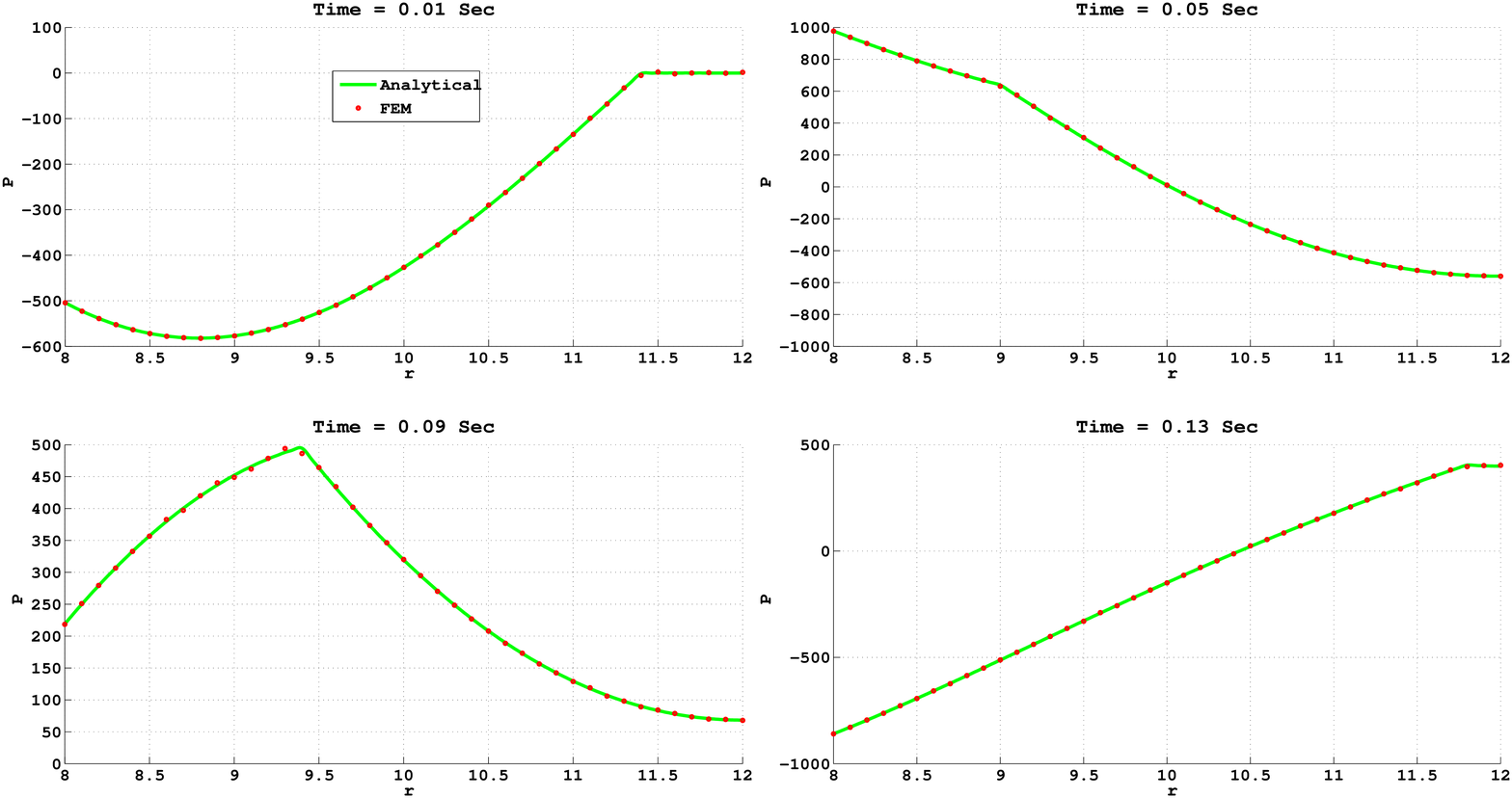}}
\caption{Pressure variation along radius for different times in the oscillating sphere (interior acoustic) problem. (a) $\theta$ = 135 degree (b) $\theta$ = 180 degree.}
\label{fig_osc_ann}
\end{figure}

Now consider the case when the oscillating sphere of radius r$_1$ is surrounded by a rigid sphere of radius r$_2$, and the acoustic fluid lies between these two spheres. 
Let $\lambda_n, n = 1,2,\cdots,\infty,$ be the positive roots of 
\begin{equation*}
\tan(r_2-r_1)x = \frac{2(r_2-r_1)x(2+r_1r_2x^2)}{4-2(r_2-r_1)^2x^2+r_1^2r_2^2x^4}.
\end{equation*}
For the case when $A=V\omega \cos\omega t$, we get (with $k\equiv \omega/a_0$)
\begin{align*}
p&=-\rho_0 a_0V\cos\theta\biggl\{
\frac{kr_1^3\cos\omega t}{r^2D}
\left[k[r(2-k^2r_2^2)-2r_2]\cos k(r_2-r)+[2-k^2r_2(r_2-2r)]\sin k(r_2-r)\right] \\
&+\sum_{n=1}^{\infty} \frac{2\lambda_nd_n\left(\frac{\lambda_n}{k}\right)\cos\lambda_na_0 t}{\left(\frac{\lambda_n}{k}\right)^2-1}\biggr\},
\end{align*}
where
\begin{align*}
D&=2k(r_2-r_1)(k^2r_1r_2+2)\cos k(r_2-r_1)-[4-2k^2(r_2-r_1)^2+k^4r_1^2r_2^2]\sin k(r_2-r_1),\\
d_n&=\frac{r_1^3}{r^2C}\left[\lambda_n(2r_2-2r+\lambda_n^2rr_2^2)\cos\lambda_n(r_2-r) - (2+2\lambda_n^2rr_2-\lambda_n^2r_2^2)\sin\lambda_n(r_2-r)\right],\\
C&=\lambda_n^3\left[(r_2-r_1)[r_1^2(\lambda_n^2r_2^2-2)-2r_1r_2-2r_2^2]\cos\lambda_n(r_2-r_1)+2\lambda_nr_1r_2(r_1^2+r_2^2)\sin\lambda_n(r_2-r_1)\right].
\end{align*}
For the numerical simulations, we take $r_1=8$ m, $r_2=12$ m, $V=2$ and $\omega=200$. A uniform mesh of $n_r\times n_{\theta}=40\times 4$ axisymmetric nine-node elements
is used to discretize the semi-circular domain, and the time step chosen is $\deltat=2.5\times 10^{-5}$ s. Figure~\ref{fig_osc_ann}
shows the excellent agreement between the analytical and the finite element solutions for two different $\theta$ values. 

In addition to the above examples, we have also simulated the problem of a piston on a spherical baffle, and found that the results match with those presented by
Pinsky and Abboud~\cite{15}.
\subsection{Structural Problems}
\subsubsection{Clamped circular plate under ring pressure load}
A circular plate of radius $a=1$ m and thickness $h=0.01$ m is clamped at its boundary. A ring pressure load is applied from $r_1=0.2$ m to $r_2=0.4$ m on top of the plate. 
Two types of transient loading cases are considered: (i) $p\sin\omega t$ with $p=2$ and $\omega=500$, (ii) $pe^{-\omega t}$ with $p=2$ and $\omega=200$. 
The initial displacement and velocity of the plate is assumed to be zero. The structural damping is assumed to be of Rayleigh type ($\bC=\alpha\bM+\beta\bK$). 
The Young Modulus ($E$), Poisson ratio ($\mu$) and density ($\rho$) of the plate material are taken as $210\times 10^9\ \text{N/m}^{2}$, 0.3 and $7800\ \text{kg/m}^{3}$, respectively.

Let $\lambda_m$, $m=1,2,\ldots,\infty$, be the infinite roots of 
\begin{equation*}
J_0(\lambda)I_1(\lambda)+I_0(\lambda)J_1(\lambda)=0,
\end{equation*}
where $J_0$ and $I_0$ represent the Bessel and modified Bessel functions of the first kind.
For the axisymmetric problem under consideration, the modes are \cite{14}
\begin{equation*} \label{circmode}
[U_{3}(r,\theta)]_m= J_{0}\left(\frac{\lambda_mr}{a}\right)-\frac{J_{0}(\lambda_m)}{I_{0}(\lambda_m)}I_{0}\left(\frac{\lambda_mr}{a}\right).
\end{equation*}
The displacement solution is given by 
\begin{equation*} \label{circtot}
u_3(r,t) = \sum_{m=0}^{\infty}[\eta(t)]_{m}[U_{3}(r)]_{m}.
\end{equation*}
Using the Laplace transformation technique, we obtain
\begin{equation*}
[\eta(t)]_m = \frac{p D_m T_m}{a\rho h S_m},
\end{equation*}
where
\begin{align*}
D_{m} &= \frac{\left[r_{2}J_{1}\left(\frac{\lambda_mr_{2}}{a}\right) - r_{1}J_{1}\left(\frac{\lambda_mr_{1}}{a}\right)\right] - 
C_{m}\left[r_{2}I_{1}\left(\frac{\lambda_mr_{2}}{a}\right) - r_{1}I_{1}\left(\frac{\lambda_mr_{1}}{a}\right)\right]}{\gamma_m\lambda_mJ_0^2(\lambda_m)} , \\
C_{m} &= \frac{J_{0}\left(\lambda_m\right)}{I_{0}\left(\lambda_m\right)} , \\
\gamma_m &= \omega_m\sqrt{1-\zeta_m^2} , \\
\zeta_m &= \frac{\alpha + \beta\omega_m^2}{2\omega_m} ,\\ 
\omega_{m} &= \frac{\lambda_m^{2}}{a^{2}}\sqrt{\frac{D}{\rho h}} , \\
D &= \frac{E h^{3}}{12(1-\mu^{2})}. 
\end{align*}

When the pressure loading is of the type $p\sin\omega t$, we obtain
\begin{align*}
T_{m} &= \gamma_m\left[(\omega_m^2-\omega^2)\sin(\omega t) - 2\zeta_m\omega_m\omega\cos(\omega t)\right]+ \\
&\quad e^{-\zeta_m\omega_mt}\left[\omega(\zeta_m^2\omega_m^2+\omega^2-\gamma_m^2)\sin(\gamma_mt)+2\zeta_m\omega_m\omega\gamma_m\cos(\gamma_mt)\right] ,\\
 S_m & = \zeta_m^4\omega_m^4 + (\omega^2-\gamma_m^2)^2 + 2\zeta_m^2\omega_m^2(\omega^2+\gamma_m^2).
\end{align*}
while when it is of the type $pe^{-\omega t}$, we get
\begin{align*}
 T_m &= e^{-\zeta_m\omega_mt}\left[(\omega-\zeta_m\omega_m)\sin(\gamma_mt)-\gamma_m\cos(\gamma_mt)\right]+\gamma_me^{-\omega t}, \\
 S_m &= (\omega - \zeta_m\omega_m)^2 + \gamma_m^2.
\end{align*} 

\begin{figure}
 \centering
\subfloat[]{\label{circ_damped_initial}\includegraphics[trim = 7cm 1cm 7cm 1cm, width = 12cm,height=7 cm]{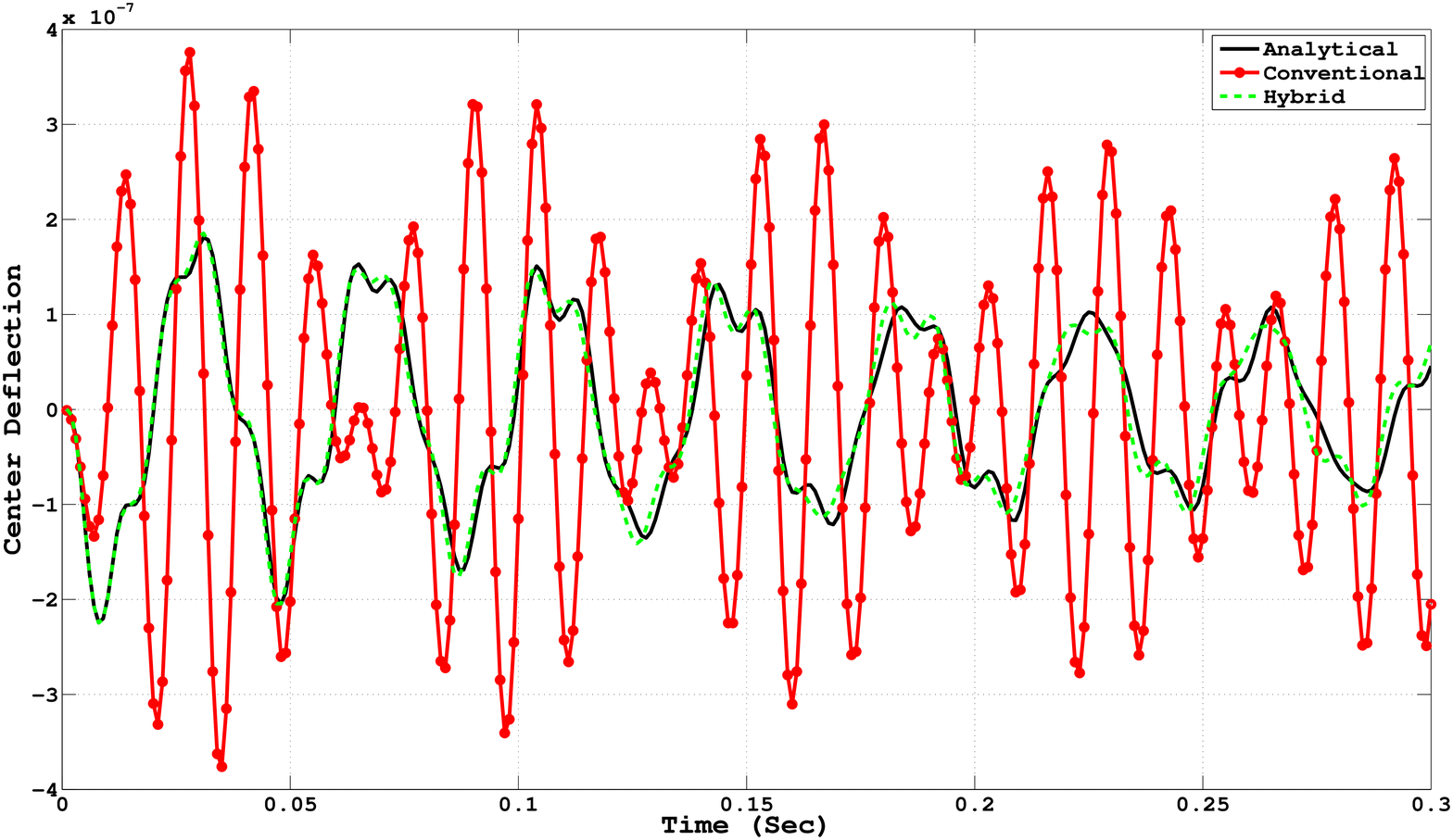}}\\
\subfloat[]{\label{circ_damped_steady}\includegraphics[trim = 7cm 1cm 7cm 1cm, width = 12cm,height=7cm]{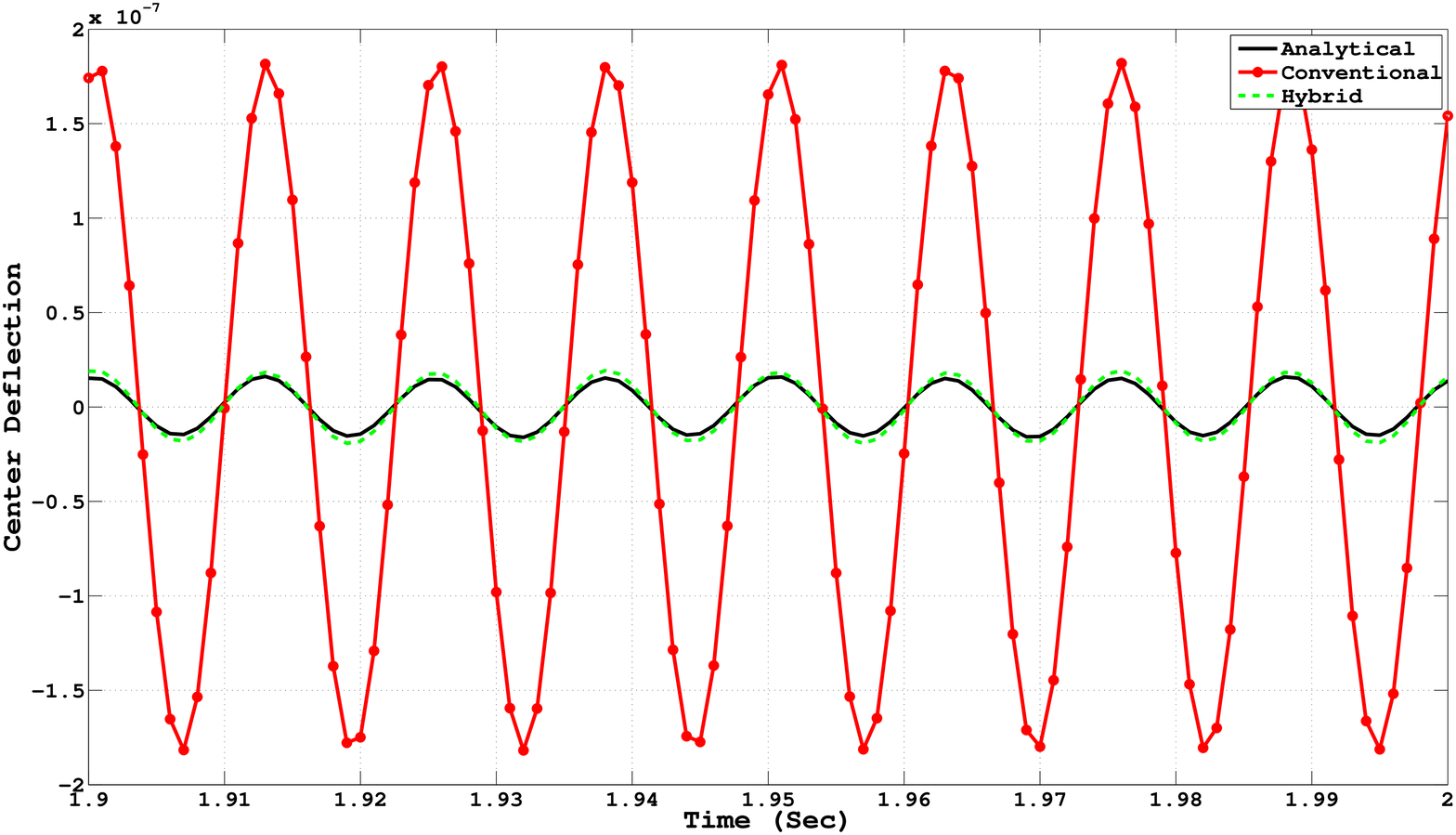}}
\caption{Center deflection for a ring pressure load of $2\sin 500 t$ in the clamped circular plate problem. (a) Initial transient response (b) Periodic steady-state response.}
\label{fig_circ_damped}
\end{figure}

For the sinusoidal loading case, we have considered damping with $\alpha = 5.517$ and $\beta = 8.62e-6$. The finite element results for the center point deflection 
obtained using a uniform mesh of $n_r\times n_z=20\times 2$ axisymmetric four-node elements with $\deltat=2\times 10^{-4}$ s are presented in Figure~\ref{fig_circ_damped}.

As can be seen, both, the initial transients and the periodic steady-state response are captured more accurately by the hybrid finite element formulation.
With the use of an equivalent $n_r\times n_z=10\times 1$ axisymmetric nine-node element mesh with $\deltat$ the same, one again finds the hybrid finite element results
to be more accurate, although the difference between the hybrid and conventional formulation results is more dramatic in the case of four-node elements.
If the difference between the analytical and the finite element results is normalized by the maximum magnitude of analytical central displacement between 0 and 2 s
given by 2.25e-7 m, then for the same mesh, the maximum errors for four-node conventional and hybrid elements are $221.27\%$ and
$18.3\%$, respectively, while for nine-node conventional and hybrid elements, the corresponding figures are $24.9\%$ and $9.14\%$, respectively. This result clearly
brings out one of the main points of this work, namely, that while there may be a need for `high-frequency dissipation' with the use of conventional elements, such a need is bypassed
using hybrid elements.

\begin{figure}
 \centering
\subfloat[]{\label{fig_circ_undamped}\includegraphics[trim = 7cm 1cm 7cm 1cm, width = 12cm,height=7.0 cm]{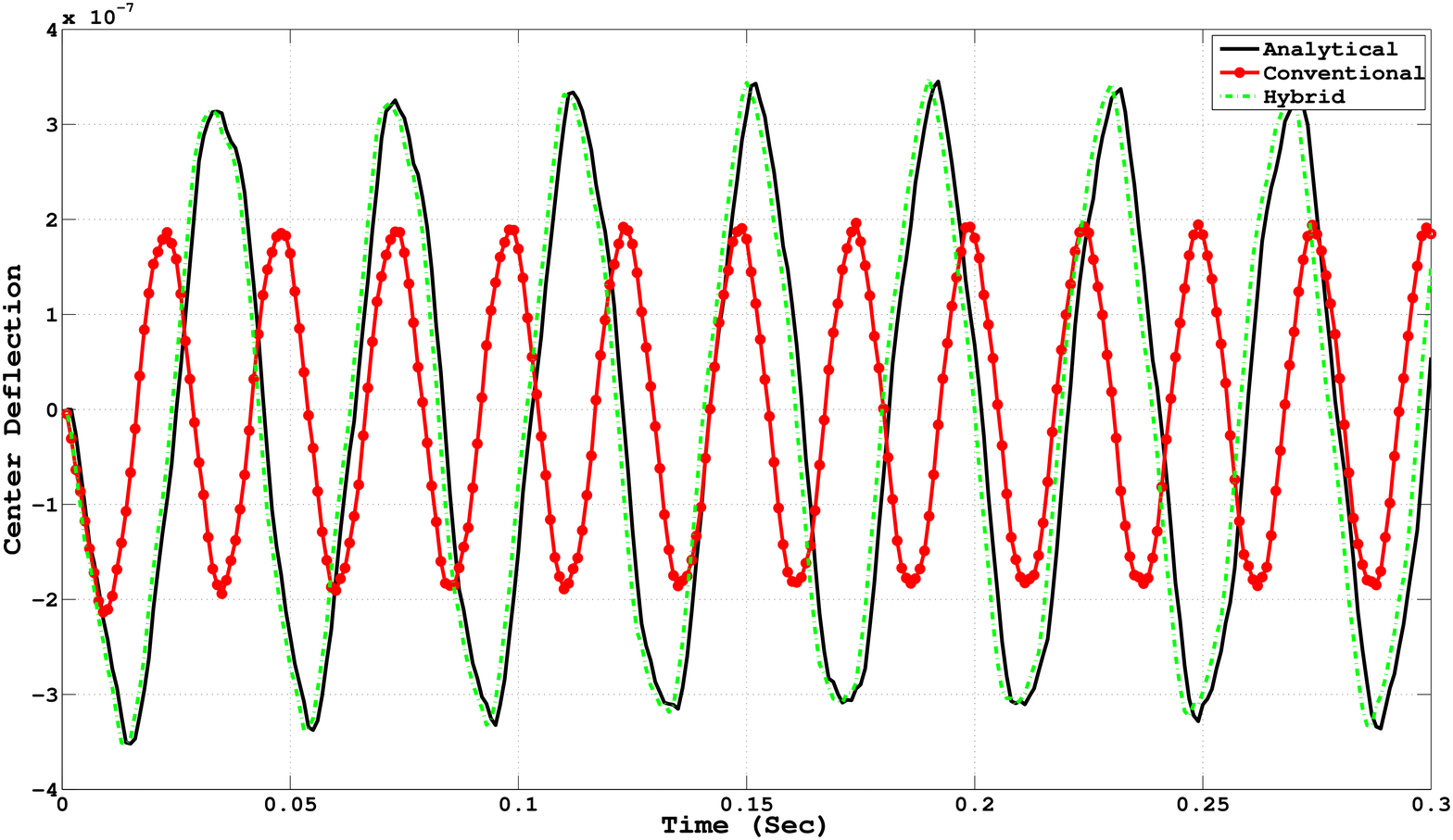}}\\
\subfloat[]{\label{fig_circ_undamped_energy}\includegraphics[trim = 7cm 1cm 7cm 1cm, width = 12cm,height=7.0cm]{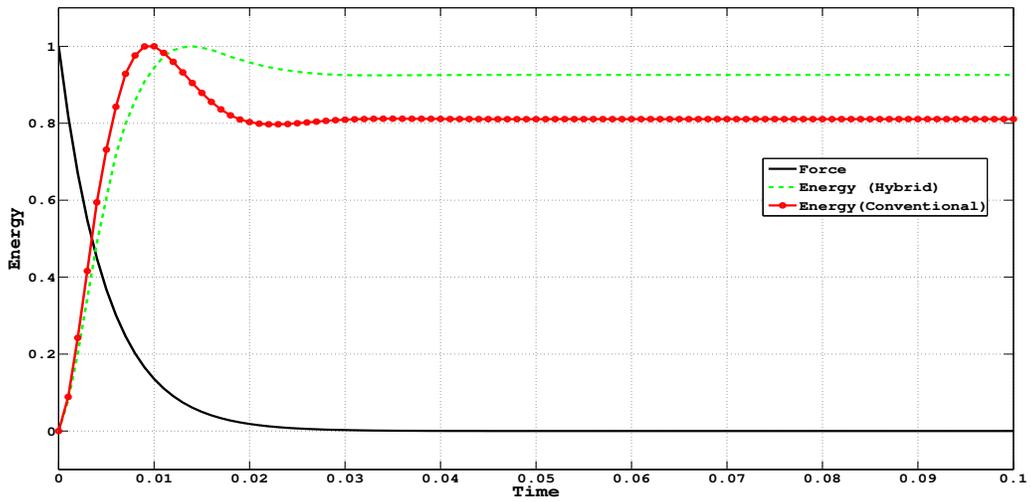}}
\caption{Force and energy variations for a ring pressure load of $2\text{e}^{\text{-200t}}$ in the clamped circular plate example. (a) Center deflection (b) Conservation of energy.}
\end{figure}

In order to show the energy conserving characteristic of the trapezoidal rule, the exponential pressure loading is considered in the undamped case ($\alpha=\beta=0$).
A uniform mesh of $n_r\times n_z= 40\times 4$ axisymmetric four-node elements is used to mesh the domain and we use $\deltat=10^{-4}$ s. Figure \ref{fig_circ_undamped} 
once again shows the almost perfect match between the analytical and hybrid finite element results, and the significant errors in the
conventional finite element formulation. Since there is a reaction traction at the clamped edge, the linear and angular momenta are not conserved. However,
since the velocity at the clamped edge is zero, the energy is conserved once the load decays to zero.
Figure~\ref{fig_circ_undamped_energy} shows this conservation property for both conventional and hybrid elements. The force and the conventional and hybrid
formulation energies are normalized with respect to their corresponding maximum values of 2, 7.863e-9 and 1.071e-8, respectively.
\subsubsection{Clamped skew plate subjected to uniform pressure load}
\begin{figure}
 \centering
\subfloat[]{\label{skw_plt}\includegraphics[width = 10cm,height=4 cm]{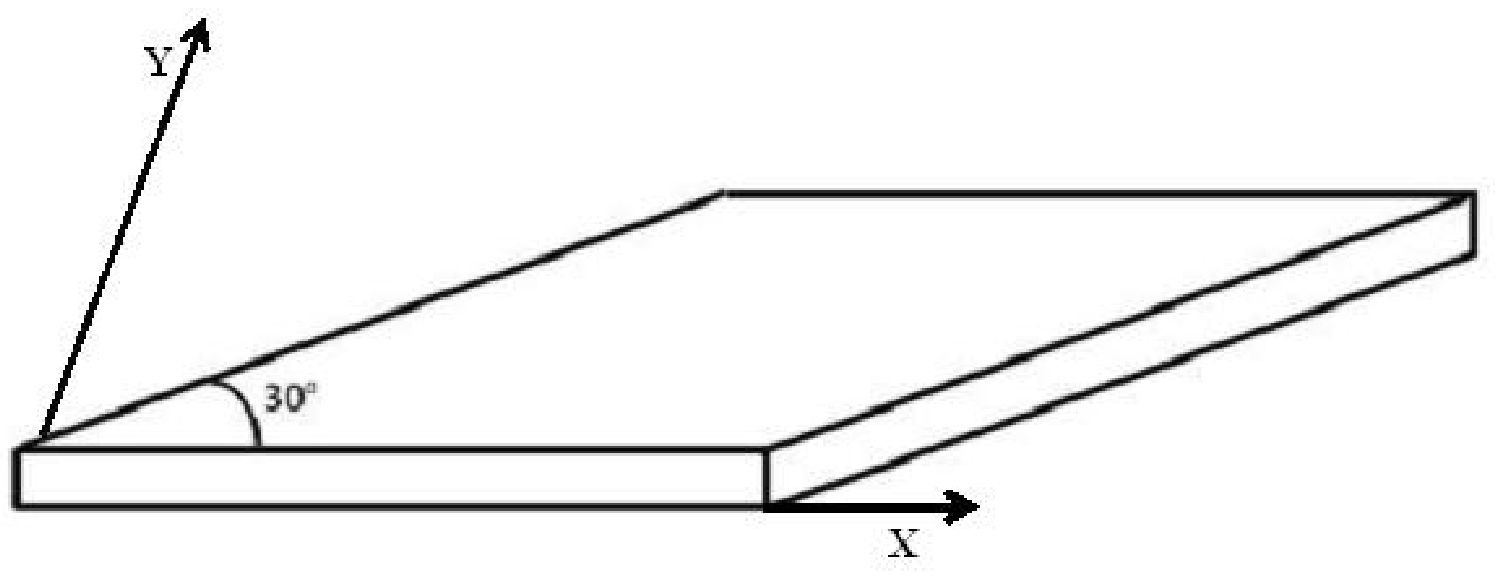}}\\
\subfloat[]{\label{fig_skw_damped}\includegraphics[trim = 7cm 1cm 7cm 1cm, width = 12cm,height=6.5cm]{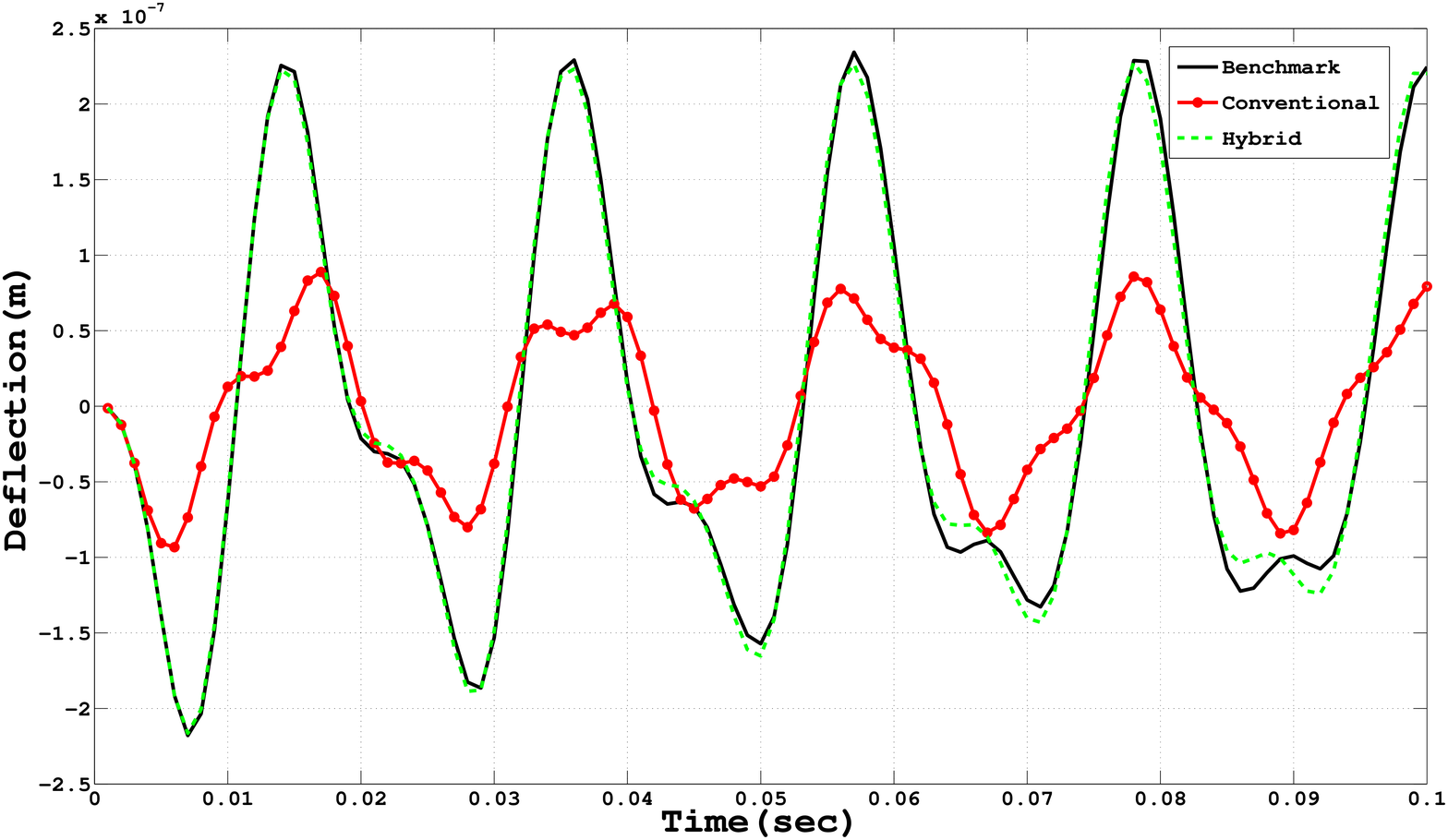}}
\caption{Skew plate problem. (a) Geometry (b) Center point deflection.}
\end{figure}

This example shows that the hybrid formulation can handle mesh distortions better than conventional element. Consider a rhomboid skew plate (see Figure~\ref{skw_plt})
of side 1 m and thickness 0.01 m clamped at all of its boundaries. The plate is subjected to uniform pressure $2\sin 300t$ on the top surface. The Young modulus, Poisson ratio 
and density of the plate are taken as $20\times10^9\ \mathrm{N/m^2}$, 0.3 and $8000\ \mathrm{kg/m^3}$, respectively. The plate damping parameters are taken as 
$\alpha = 5.517$ and $\beta = 8.62\times 10^{-6}$. Since there is no analytical solution available, we use a converged solution obtained with a fine mesh of $80\times 80$ 
twenty-seven node hexahedral elements and $\deltat = 2.5\times 10^{-5}$ s as the benchmark solution for comparing coarse-mesh solutions obtained using the hybrid and 
conventional formulations. The solution for the center-point ($(x_c,y_c)=(0.933,0.25)$) deflection obtained using a $40\times 40\times 2$ mesh of hybrid and conventional 
8-node hexahedral elements, and 
$\deltat = 10^{-4}$ s is plotted against the benchmark solution as shown in Figure~\ref{fig_skw_damped}, once again showing the superior performance of the hybrid formulation.
Taking the maximum center deflection $2.34\times 10^{-7}$ in the time interval $[0,0.1]$~s of the benchmark solution as the normalizing factor, for the 8-node mesh the
maximum errors over the time interval $[0,0.1]$ s are $79.5\%$ and $8.4\%$ for conventional and hybrid elements, respectively, while for an equivalent 27-node $20\times20\times 1$ 
mesh with $\deltat = 10^{-4}$ s, the errors are $21.55\%$ and $3.58\%$, respectively.
\subsubsection{Plate subjected to exponentially decaying loads}
\begin{figure}
 \centering
\subfloat[]{\label{con_linmom}\includegraphics[trim = 7cm 1cm 7cm 1cm,width = 12cm,height=6.5 cm]{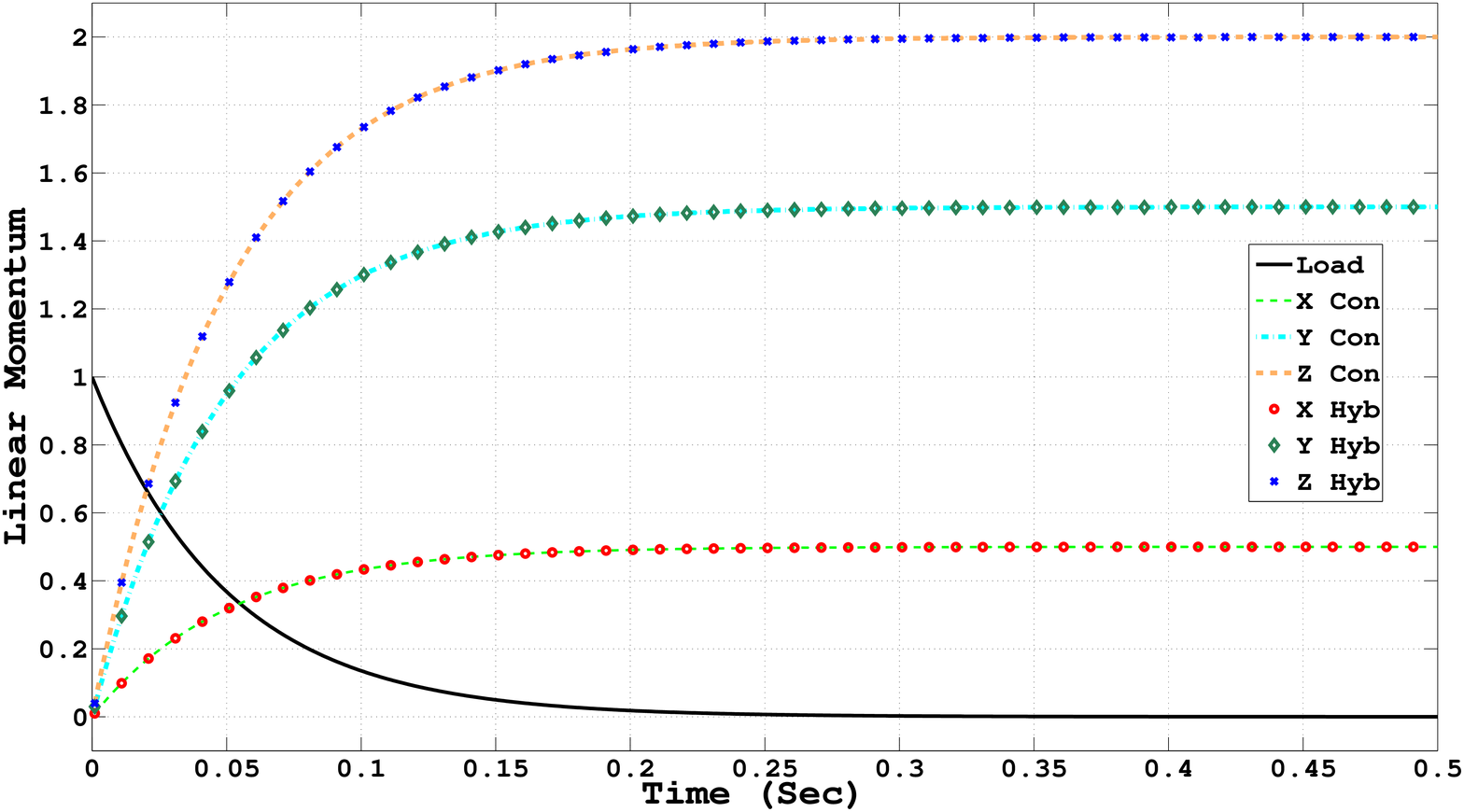}}\\
\subfloat[]{\label{con_angmom}\includegraphics[trim = 7cm 1cm 7cm 1cm,width = 12cm,height=6.5cm]{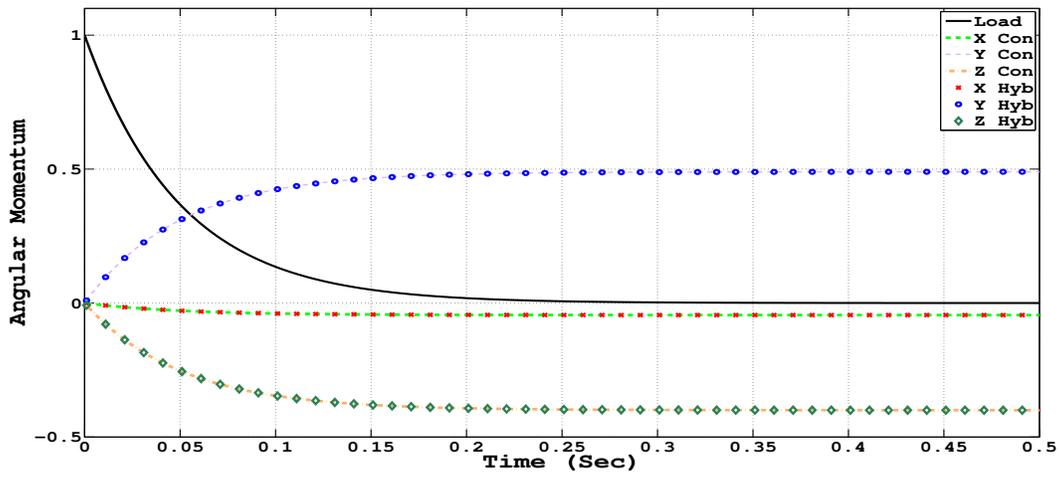}}
\caption{Variation of force and linear and angular momenta as a function of time; `Con' and `Hyb' denote conventional and hybrid formulations, respectively.} \label{conserve_momentum}
\end{figure}

\begin{figure}
\begin{center}
\includegraphics[trim = 7cm 1cm 7cm 1cm, width = 12cm, height = 6.5cm]{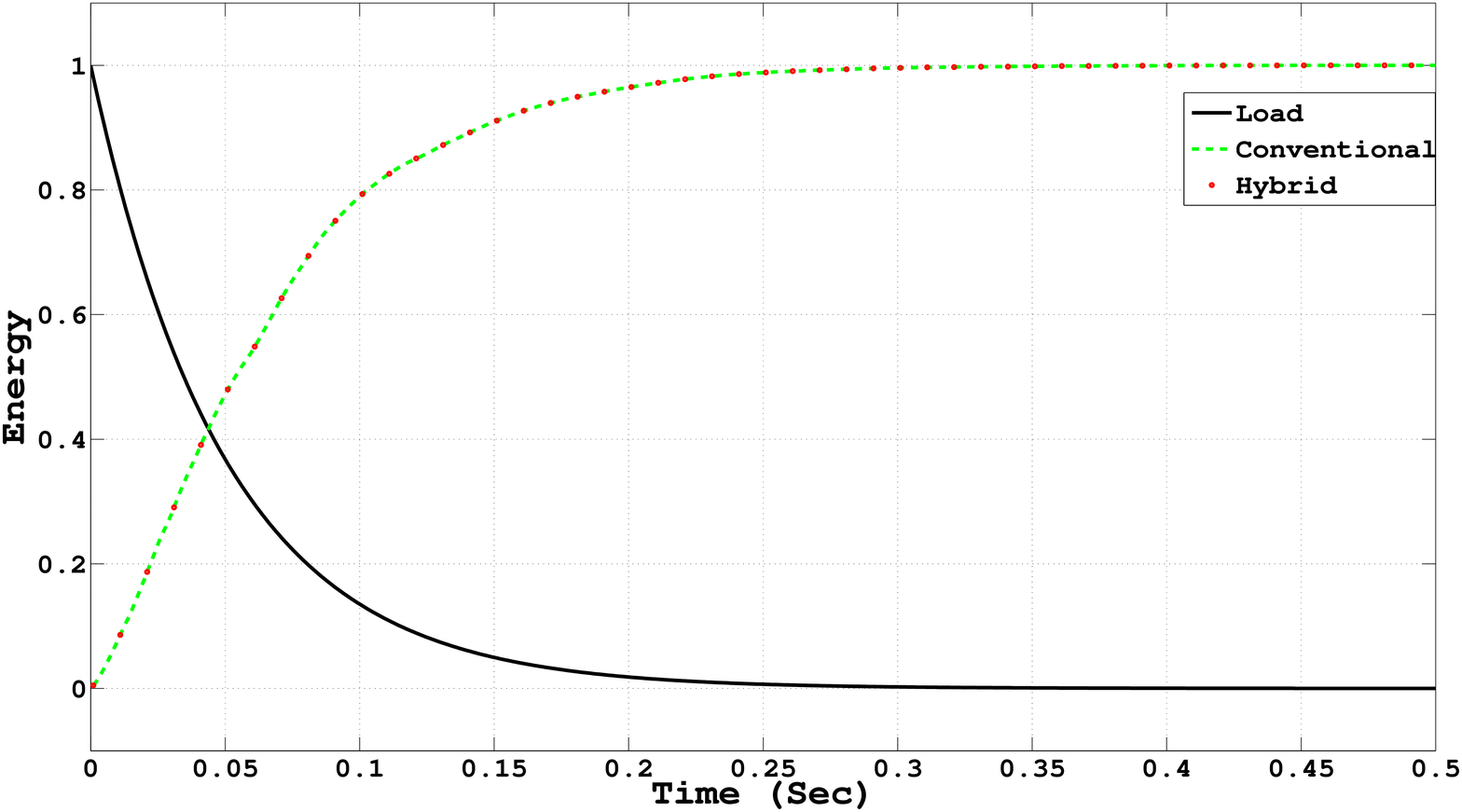}
\caption{Variation of energy with time in the plate subjected to exponentially decaying loads example.} \label{conserve_energy}
\end{center}
\end{figure}

The purpose of this example is to show the energy-momentum conserving properties of the trapezoidal rule.
Consider a plate of dimension $1 \text{ m}\times0.1 \text{ m}\times0.01\text{ m}$ with edges along the co-ordinate axes, and whose boundaries are traction free. 
The two opposite corners (0,0,0) and (1.0,0.1,0.01) are subjected to nodal forces $(F_x,F_y,F_z)=(30,40,50)e^{-20 t}$ and $(-20,-10,-10)e^{-20 t}$, respectively.
The Young modulus, Poisson ratio and density of the plate material are taken as $210\times10^7\ \mathrm{N/m^2}$, 0.3 and $7800\ \mathrm{kg/m^3}$, respectively. The
plate damping parameters $\alpha$ and $\beta$ are taken as zero. A mesh of $20\times1\times2$ 27-node hexahedral elements and time step $\deltat=10^{-4}$ s is used.
Figure~\ref{conserve_momentum} shows the variation of the linear and angular momenta (for each of x, y and z components), respectively.
The force variation has also been plotted to show that the linear and angular momenta are conserved once the force decays to zero.
The energy variation (normalized using the maximum value of 4.94) with time is shown in Figure~\ref{conserve_energy}.
\subsection{Coupled Problems}
\subsubsection{Spherical shell in exterior acoustic domain}
\begin{figure}
\begin{center}
\includegraphics[trim = 7cm 1cm 7cm 1cm, width = 12cm, height = 6.5cm]{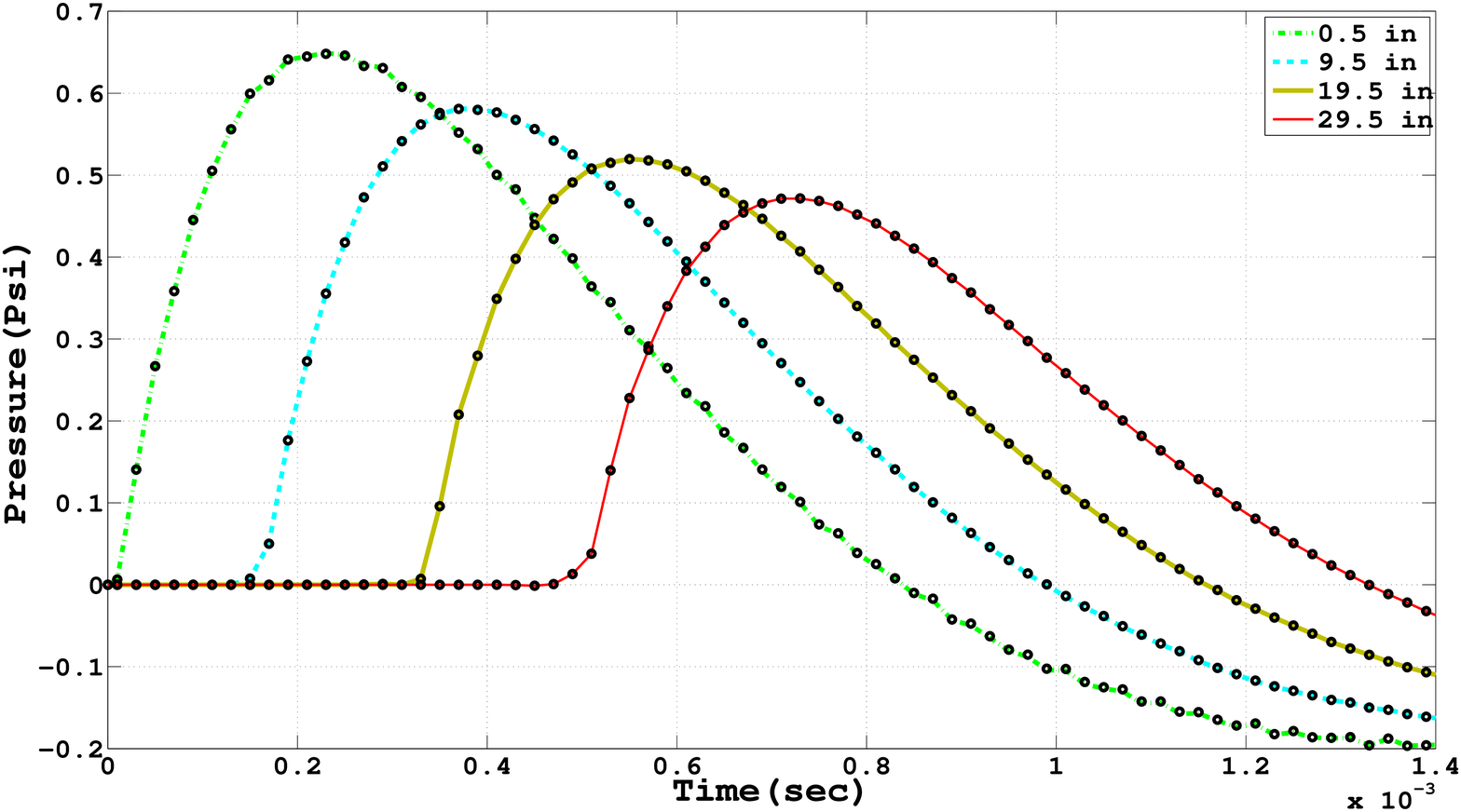}
\caption{Pressure variation with time at different distances from the shell; solid lines and black dots represent the conventional and hybrid results, respectively.} \label{Akkas_fig5}
\end{center}
\end{figure}

The response of a spherical shell immersed in a heavy acoustic fluid and subjected to an impulsive pressure load has been analyzed by Akkas~\cite{12}. By comparing
our results shown in Figure~\ref{Akkas_fig5} against their Figure~5, we see that almost identical results are obtained using the trapezoidal rule.
We have terminated the acoustic domain at a radius of 106 in. An axisymmetric 9 node quarter circular mesh (1 element along shell thickness, 30 elements in radial direction of the
acoustic domain and 4 elements along the $\theta$ direction) with time step $\deltat = 5\times 10^{-6}$ s is used. The results obtained using conventional and hybrid elements are 
identical in this case since the loading is spherically symmetric so that there is no shearing.
\subsubsection{Cylindrical shell with interior acoustic fluid}
\begin{figure}
\begin{center}
\includegraphics[trim = 3cm 1cm 8cm 1cm, width = 3.5cm, height = 5.5cm]{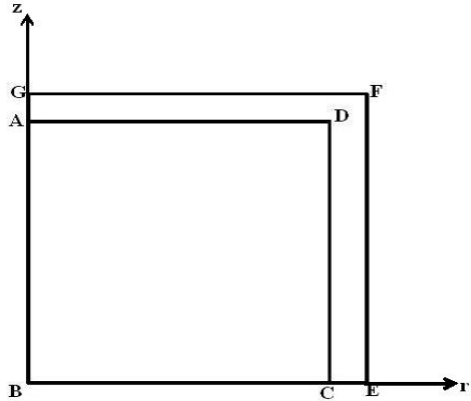} \\[3mm]
\caption{Domain used for meshing in the coupled cylindrical shell problem.} \label{coupled_cyl}
\end{center}
\end{figure}

\begin{figure}
\begin{center}
\includegraphics[trim = 7cm 1cm 7cm 1cm, width = 14cm, height = 8cm]{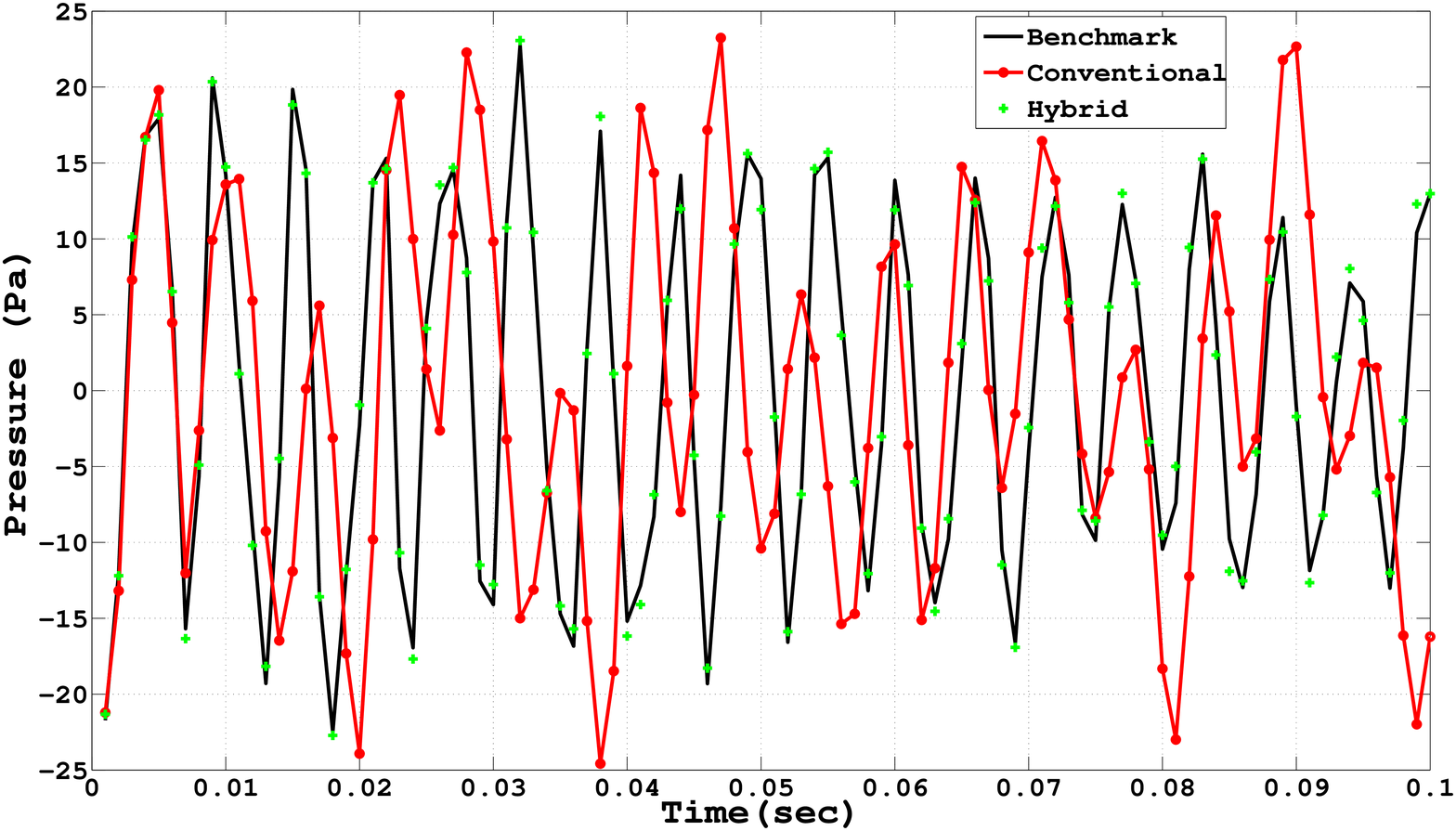}
\caption{Variation of pressure at $(r,z)-(0.5,0.5)$ with time in the coupled cylindrical shell problem.} \label{fig_coupled}
\end{center}
\end{figure}

Consider a cylindrical shell of outer radius 1.01 m, maximum height 2.02 m and uniform thickness 0.01 m. The acoustic fluid lies inside the cylinder. A normal sinusoidal ring line 
load is applied on the outside surface at $z=0$. The Young modulus, Poisson ratio and density for the shell material are 
$210\times 10^{9}\ \mathrm{N/m^2}$, $0.3$ and $7800\ \mathrm{kg/m^3}$. In order to create a significant coupling effect, we take the acoustic fluid to be `heavy' with 
density $1000\ \mathrm{kg/m^3}$ and sound speed $1500\ \mathrm{m/s}$. Since the problem is axisymmetric and also symmetric about $z=0$, we have meshed the 
domain shown in Figure~\ref{coupled_cyl} using axisymmetric elements.
Taking symmetry into account, a load of $100\sin 300t$ is applied at point E. Since there is no analytical solution for this coupled problem, we take the converged solution 
obtained using a very fine nine-node mesh ($80\times 80$ in the acoustic domain and 20 elements along the shell thickness with $\deltat = 1.25\times 10^{-6}$ s) as the
benchmark solution. The pressure variation at $(r,z)=(0.5,0.5)$ using a relatively coarse $40\times 40$ mesh of axisymmetric four-node elements for the acoustic domain
with 10 elements along the shell thickness and $\deltat=2.5\times 10^{-6}$ s is shown in Figure~\ref{fig_coupled} along with the benchmark solution, demonstrating the 
good performance of hybrid elements in a coupled problem.
\section{Conclusion} \label{secconcl}
We have shown, both analytically and numerically, that the trapezoidal rule is an energy-momentum conserving algorithm within the context of linear elastodynamics
(for both conventional and hybrid elements), and an `energy-like-measure' conserving algorithm in the context of transient acoustics. Thus, from an energy-perspective, 
it is an unconditionally stable algorithm in both cases. In the presence of damping, it mimics the energy-decaying characteristics of the continuum solution in 
both linear elastodynamics and acoustics.
Thus, there is a strong justification for using this algorithm within these contexts. Problems that arise with the use of this algorithm are due to a \emph{poorly approximated
(overstiff) stiffness matrix}, and not due to any inherent shortcoming in this algorithm. Indeed, we have shown an excellent match between analytical and numerical results obtained using
a stiffness matrix constructed using hybrid elements and a trapezoidal rule for time-stepping for acoustical, structural and coupled problems. The need for `high-frequency
dissipation' is, thus, alleviated if not bypassed with the use of a hybrid formulation.
%% appendix subsections are then done as normal sections

%% \subsection{}

%% References


\bibliographystyle{model1-num-names}
\bibliography{references}






% \end{thebibliography}
%\begin{appendices}
%  \chapter{Consectetur adipiscing elit} \label{A}
%dsgsg
%  \chapter{Mauris euismod}
%sggssg
%\end{appendices}

\end{document}